\pdfoutput=1

\documentclass[11pt,twoside,a4paper,cmspaper,final,collab]{cms-tdr}
\def\svnVersion{166413}\def\cmsCernNo{2012-045}\def\cmsCernDate{\today}\def\cmsMessage{Submitted to the Journal of High Energy Physics}

\begin{document}\cmsNoteHeader{EXO-11-071}

\hyphenation{had-ron-i-za-tion}
\hyphenation{cal-or-i-me-ter}
\hyphenation{de-vices}

\RCS$Revision: 107670 $
\RCS$HeadURL: svn+ssh://svn.cern.ch/reps/tdr2/papers/EXO-11-071/trunk/EXO-11-071.tex $
\RCS$Id: EXO-11-071.tex 107670 2012-02-28 18:39:48Z alverson $
\cmsNoteHeader{EXO-11-071} 
\title{Search for microscopic black holes in pp collisions at $\sqrt{s} = 7$\TeV}

\newcommand{\BlackMax}{\textsc{BlackMax}\xspace}
\newcommand{\Charybdis}{\textsc{charybdis}\xspace}
\newcommand{\ST}{\ensuremath{S_{\mathrm{T}}}\xspace}
\renewcommand{\MD}{\ensuremath{{M_D}}\xspace}
\newcommand\mbh{\mbox{$M_\mathrm{BH}$}\ }     
\newcommand\MBH{\mbox{$M_\mathrm{BH}$}}         

\date{\today}

\abstract{
A search for microscopic black holes in pp collisions at a center-of-mass energy of 7\TeV is presented. The data sample corresponds to an integrated luminosity of 4.7\fbinv recorded by the CMS experiment at the LHC in 2011. Events with large total transverse energy have been analyzed for the presence of multiple energetic jets, leptons, and photons, which are typical signals of evaporating semiclassical and quantum black holes, and string balls. Agreement with the expected standard model backgrounds, which are dominated by QCD multijet production, has been observed for various combined multiplicities of jets and other reconstructed objects in the final state. Model-independent limits are set on new physics processes producing high-multiplicity, energetic final states. In addition, new model-specific indicative limits are set excluding semiclassical and quantum black holes with masses below 3.8 to 5.3\TeV and string balls with masses below 4.6 to 4.8\TeV. The analysis has a substantially increased sensitivity compared to previous searches.}

\hypersetup{%
pdfauthor={CMS Collaboration},%
pdftitle={Search for microscopic black holes in pp collisions at sqrt(s) = 7 TeV},%
pdfsubject={CMS},%
pdfkeywords={CMS, physics, black holes, extra dimensions}}

\maketitle 

\section{Introduction}

One of the most spectacular predictions of theories with low-scale quantum gravity is the possibility of microscopic black hole (BH) production in proton-proton collisions at the high energies offered by the Large Hadron Collider (LHC)~\cite{dl,gt}. Such models are motivated mainly by the puzzling large difference between the electroweak scale (${\sim}0.1$\TeV) and the Planck scale ($\Mpl\sim10^{16}$\TeV), known as the hierarchy problem. In this analysis, we focus on black hole production in a model with $n$ large, flat, extra spatial dimensions (ADD model) \cite{add,add1}. In this and in other models, the fundamental scale of new physics in $n$ extra dimensions is given in terms of a multidimensional Planck scale \MD, such that $M_D^{n+2} \propto M_\mathrm{Pl}^{2} R^{-n}$, where $R$ is the size of extra dimensions. Some of the conclusions also apply to black holes in the Randall--Sundrum model~\cite{RS,RS1}, with a single warped extra dimension.

This analysis extends a previous search~\cite{CMSBH} for short-lived microscopic black holes carried out by the Compact Muon Solenoid (CMS) Collaboration in $2010$. The present search is based on the full $2011$ data sample, which corresponds to an integrated luminosity of $4.7\pm0.2\fbinv$~\cite{lumi10,lumi} at a center-of-mass energy of $7$\TeV. The details of the analysis method and the underlying theory, as well as the detailed description of black hole evaporation models, can be found in the original publication~\cite{CMSBH}. Typically, microscopic black holes are characterized by the production of a large number of energetic final-state particles, $75\%$ of which are jets. Searches for black holes have also been performed by the ATLAS Collaboration~\cite{ATLASbh,ATLASdijet}.

We present our results in terms of model-independent limits on the cross section times the branching fraction into a multiparticle final state. Further, we interpret the results in terms of a set of benchmark black hole models. The analysis also extends the previous CMS search for semiclassical black holes to probe for other quantum gravity objects such as string balls~\cite{SB} and quantum black holes~\cite{MR}. It is commonly accepted that the semiclassical approximation is valid when the black hole mass is some 3--5 times larger than the \MD. The string balls are hypothetical precursors of semiclassical black holes in an extreme quantum limit, when the mass of the object is close to the Planck scale. In cases where the semiclassical approximation no longer holds, string balls may offer a more realistic description of black hole formation and decay. String balls are described by the string scale $M_S$ and string coupling $g_S$. These objects would evaporate similarly to black holes, except that the evaporation would occur at the Hagedorn temperature, which does not depend on the string-ball mass~\cite{SB,Hagedorn}, unlike the Hawking temperature~\cite{Hawking}, which decreases as the black hole mass increases. Another possibility is that a light black hole with mass close to the Planck scale may evaporate faster than it thermalizes, resulting predominantly in a nonthermal decay into a pair of jets, rather than into high-multiplicity final states~\cite{MR,Calmet,qbh1}. We search for production of these objects, referred to as quantum black holes, and for their decay in both the ADD scenario and in the Randall--Sundrum model of low-scale gravity with a single ($n = 1$) compact extra dimension.

\section{The CMS detector}

A detailed description of the CMS experiment can be found elsewhere~\cite{CMS}. The CMS detector consists of a 3.8\unit{T} superconducting solenoid enclosing a silicon pixel and strip tracker, a crystal electromagnetic calorimeter (ECAL), and a brass-scintillator hadronic calorimeter (HCAL). The finely segmented ECAL employs lead-tungstate crystals with transverse dimensions: $\Delta \eta \times \Delta \phi = 0.0174\times0.0174$. The HCAL cells are grouped in projective towers, of granularity $\Delta \eta \times \Delta \phi = 0.087\times0.087$ at central rapidities and increasing progressively in the forward region. Here, $\phi$ and $\theta$ are the azimuthal and polar angles, with $\theta$ measured with respect to the direction of the counterclockwise beam. The pseudorapidity $\eta$ is defined as $-\ln[\tan(\theta/2)]$. Muons are measured in the pseudorapidity range $|\eta|< 2.4$ in gas-ionization detectors embedded in the steel return yoke.

The CMS trigger system, used to select the most interesting events, consists of two levels. A first level (L1), composed of custom electronics, uses information from the calorimeters and muon detectors to decrease the event rate to 80\unit{kHz}. A software-based High Level Trigger (HLT) further decreases the event rate to 350--400\unit{Hz} for data storage.

As in the previous analysis, we use data collected with a suite of \HT triggers, where \HT is defined as a scalar sum of the transverse energies (\ET) of the jets above a threshold.\footnote{Energetic electrons and photons are also reconstructed as jets at the trigger level.} There have been changes introduced in the trigger logic both at L1 and HLT since 2010. We now use jets corrected for the calorimeter response to calculate the \HT variable at the HLT (uncorrected jets are still used at L1). Also, the minimum \HT thresholds, as well as the minimum jet \ET requirement for a jet to be counted towards \HT, have been increased to account for pileup effects and to allow for the increased instantaneous luminosity of the LHC. This minimum jet \ET threshold is $10$\GeV at L1, and $40$\GeV at the HLT. The minimum \HT threshold at the HLT varies between $350$ and $650$\GeV, depending on the instantaneous luminosity. Only jets reconstructed at central pseudorapidities $|\eta| < 3.0$ are used in the \HT calculations at L1 and at the HLT for the 2011 data-taking period. The trigger is fully efficient for events with \HT above $800$\GeV. In order to explore all possible black hole decay modes, the entire analysis was also repeated using data collected with multimuon or missing transverse energy (\MET) triggers, but this yielded no events consistent with the expected black hole production.

\section{Event reconstruction and Monte Carlo samples}

Jets are reconstructed offline using energy deposits in the HCAL and ECAL, which are clustered using an infrared-safe anti-$k_T$ algorithm with a distance parameter of $0.5$~\cite{anti-kt,fastjet,fastjet_code}. Quality criteria are applied to jets in order to remove calorimeter noise and noncollision background~\cite{JES}. We require jets to have $\ET > 20$\GeV and to be reconstructed within $|\eta| <  2.6$. Further, jet energies are corrected for calorimeter nonuniform response with correction factors derived from Monte Carlo (MC) simulation and dijet events from collision data~\cite{JES}. The transverse energy resolution for jets $\Delta \ET/\ET$ is better than 15\% in the range considered. We reconstruct \MET~as the modulus of the negative vector sum of transverse energies in the individual calorimeter towers and is further corrected for the jet energy scale and for muon momenta measured in the trackers~\cite{MET}.

Electrons and photons are reconstructed as isolated energy deposits in the ECAL with a shape consistent with that expected for electromagnetic showers. Electrons are required to have a track matched to the calorimeter cluster, while photons are required to have no matching hits in the silicon layers. Electrons and photons are selected with $\ET > 20$\GeV and are required to be reconstructed in the fiducial barrel ($|\eta| < 1.44$) or endcap ($1.56 < |\eta| < 2.4$) regions. The ECAL has excellent energy resolution, for example contributing $1$\GeV to the observed width of the Z boson using \Pep\Pem\ pairs with low bremsstrahlung loss, measured in the barrel region~\cite{ECAL}.

Muons are reconstructed as matched tracks in the muon spectrometer and the silicon tracker. Muons are selected with $|\eta| < 2.1$ and $\pt > 20$\GeV, and are required to be isolated from other tracks. Requiring the muons to have distance of the closest approach $< 0.2$\unit{cm} helps to suppress backgrounds from cosmic-ray muons. Here, the distance of the closest approach is defined as the shortest distance between the beam line and the direction of an object in the transverse plane. Performing a combined fit to the track segments measured in the silicon tracker and the muon system results in a transverse momentum resolution between $1\%$ and $5\%$ for \pt values up to $1$\TeV. The minimum separation between any two objects (jet, lepton, or photon) is required to be $\Delta R = \sqrt{(\Delta \phi)^2 + (\Delta \eta)^2} > 0.5$.

Simulated samples of semiclassical black hole events are produced using a parton-level \BlackMax~\cite{BlackMax} (v2.01) and \Charybdis~\cite{CHARYBDIS,CHARYBDIS2} (\Charybdis 2, v1.0.3) MC generators, followed by the parton showering simulation with \PYTHIA~\cite{PYTHIA} (v6.420), and a fast parametric simulation of the CMS detector~\cite{FastSim1}. The MSTW2008lo68cl~\cite{MSTW} parton distribution functions (PDF) are used for generating the signal samples. The \BlackMax and \Charybdis generators calculate the total cross section $\sigma$ from geometric considerations, assuming that $\sigma \propto \pi r_{S}^{2}$, where $r_S$ is the Schwarzschild radius~\cite{dl, gt}. The \BlackMax generator uses the approximation of rotating black holes, which includes an additional factor that depends on the number of extra dimensions $n$. The \Charybdis generator incorporates a more detailed model based on Yoshino--Rychkov corrections~\cite{YR,YR1} to the pure geometrical cross section, resulting in production cross sections that are lower by a factor of $1.36, 1.59$, and $1.78$ compared to those from \BlackMax for $n = 2$, 4, and 6, respectively. These scale factors can be used to interpret the results obtained with one framework in terms of the other.

Certain models are supported by both generators: rotating and nonrotating black holes, and black holes with mass and angular momentum loss prior to evaporation. This loss is set to be $10\%$ in the \BlackMax generator, while it is estimated in the Yoshino--Rychkov model and varies from $18\%$ to $30\%$ for $n = 2$ to $6$ in the \Charybdis generator. In addition, we use \Charybdis to simulate black hole evaporation resulting in a stable remnant with mass equal to the multidimensional Planck scale \MD, or a boiling remnant (unique to the \Charybdis generator). Both of these scenarios represent alternative descriptions of the final stage of the black hole evolution. In the case of a stable remnant, the terminal stage of a black hole is a noninteracting remnant with a mass of order \MD; in the case of a boiling remnant, a black hole undergoes a transformation into a string ball at a mass close to \MD with subsequent evaporation at a fixed temperature. We produce a number of string-ball samples using the \BlackMax generator. Finally, the \textsc{qbh} (version 1.03) matrix-element generator~\cite{qbh_gen} with CTEQ6L PDF set~\cite{CTEQ} is used, followed by the parton showering simulation with \PYTHIA and fast simulation of the CMS detector, to produce quantum black hole samples.
Table~\ref{tab:mcsamples} summarizes the models used in this search.

\begin{table}[htbp]
  \begin{center}
    \caption{Signal Monte Carlo samples and generators used in the analysis.}
    \label{tab:mcsamples}
    \begin{tabular}{|l|c|c|c|} \hline
       Sample description & \BlackMax & \Charybdis & \textsc{qbh}\\ \hline
       nonrotating BH & YES & YES & NO\\
       Rotating BH & YES & YES & NO\\
       Rotating BH with mass and & & & \\
       ~~angular momentum loss & YES ($10\%$ loss) & YES ($18-30$\% loss) & NO\\
       Rotating BH, low multiplicity regime & NO & YES & NO\\
       Boiling remnant & NO & YES & NO\\
       Stable remnant & NO & YES & NO\\\hline
       String balls & YES & NO & NO\\ \hline
       Quantum BH & NO & NO & YES \\
       \hline
    \end{tabular}
  \end{center}
\end{table}

\section{Analysis method}

The total transverse energy is used to separate black hole candidate events from backgrounds. A variable \ST is defined as a scalar sum of the transverse energy (\ET) of individual objects: jets, electrons, photons, and muons passing the selections described above. Only objects with $\ET > 50$\GeV enter into the sum for the calculation of \ST and count towards the final-state multiplicity $N$. This rather high minimum transverse energy requirement makes the analysis insensitive to the jets from pileup and reduces the SM backgrounds by a few orders of magnitude, while being fully efficient for black hole decays. We further add the measured \MET\ in the event to the \ST, if $\MET > 50$\GeV. Generalization of the \ST definition to include \MET\ is important for testing black hole models with a significant amount of missing energy such as models with a stable noninteracting remnant. A spurious \MET\ may arise in an event as the result of mismeasurement of the jets. However, we have checked that the consequent effect on \ST of double counting energy in both the jet and \MET\ contributions is negligible. Note that by construction, particle misidentification does not affect the total transverse energy in the event considerably.

Depending on the details of the black hole evaporation, a large variation of particle multiplicity in the final state, and a large range of missing transverse energies are possible. While resulting in quite different signatures, these variations typically have very little effect on the value of \ST in the event. A recent work on quantum-gravity black holes~\cite{qcbh1} also suggests a larger number of softer particles produced in black hole evaporation than in the semiclassical case, further emphasizing the importance of \ST as a largely model-independent variable for black hole searches.

The main background to black hole production arises from QCD multijet production, which dominates the event rates at large \ST. Smaller backgrounds come from $\gamma$/W/Z+jets and \ttbar production. These smaller backgrounds are negligible at large values of \ST and contribute less than 1\% to the total background after the final selection. We estimate their contribution from MC simulation, using the \MADGRAPH~\cite{MadGraph} leading-order parton-level event generator (with up to three extra partons included in the simulation) with the CTEQ6L PDF set followed by \PYTHIA~\cite{PYTHIA} parton showering and full CMS detector simulation via \GEANTfour~\cite{GEANT4}. For the dominant QCD background, however, we estimate backgrounds from the observed data using the \ST multiplicity invariance method~\cite{CMSBH}. This method relies on the independence of the shape of the \ST spectrum on the number of final-state objects $N$; an empirical observation extensively checked by using various MC samples (\ALPGEN and \PYTHIA) as well as low-multiplicity data. The origin of this invariance lies in the collinear nature of the final-state radiation, which typically does not change the total transverse energy in the event; hence the independence of the \ST spectrum of the jet multiplicity for the QCD background. This invariance allows us to predict the shape of the \ST spectrum for any number of objects using the dijet data, which has been studied extensively for presence of new physics in dedicated analyses~\cite{dijets1,dijets2,dijets3}.

\begin{figure}[htbp]
\centering
\includegraphics[width=0.49\textwidth]{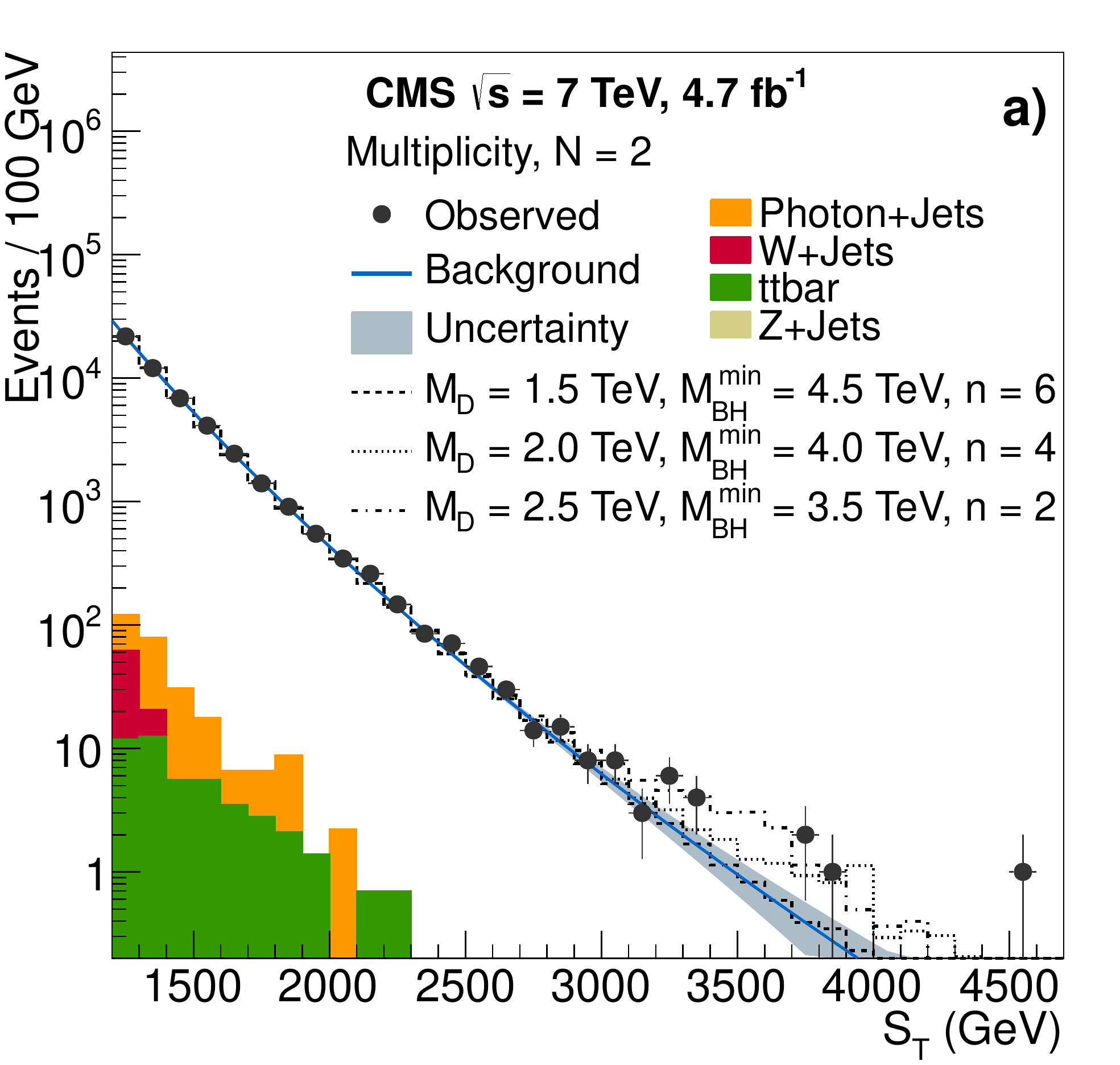}
\includegraphics[width=0.49\textwidth]{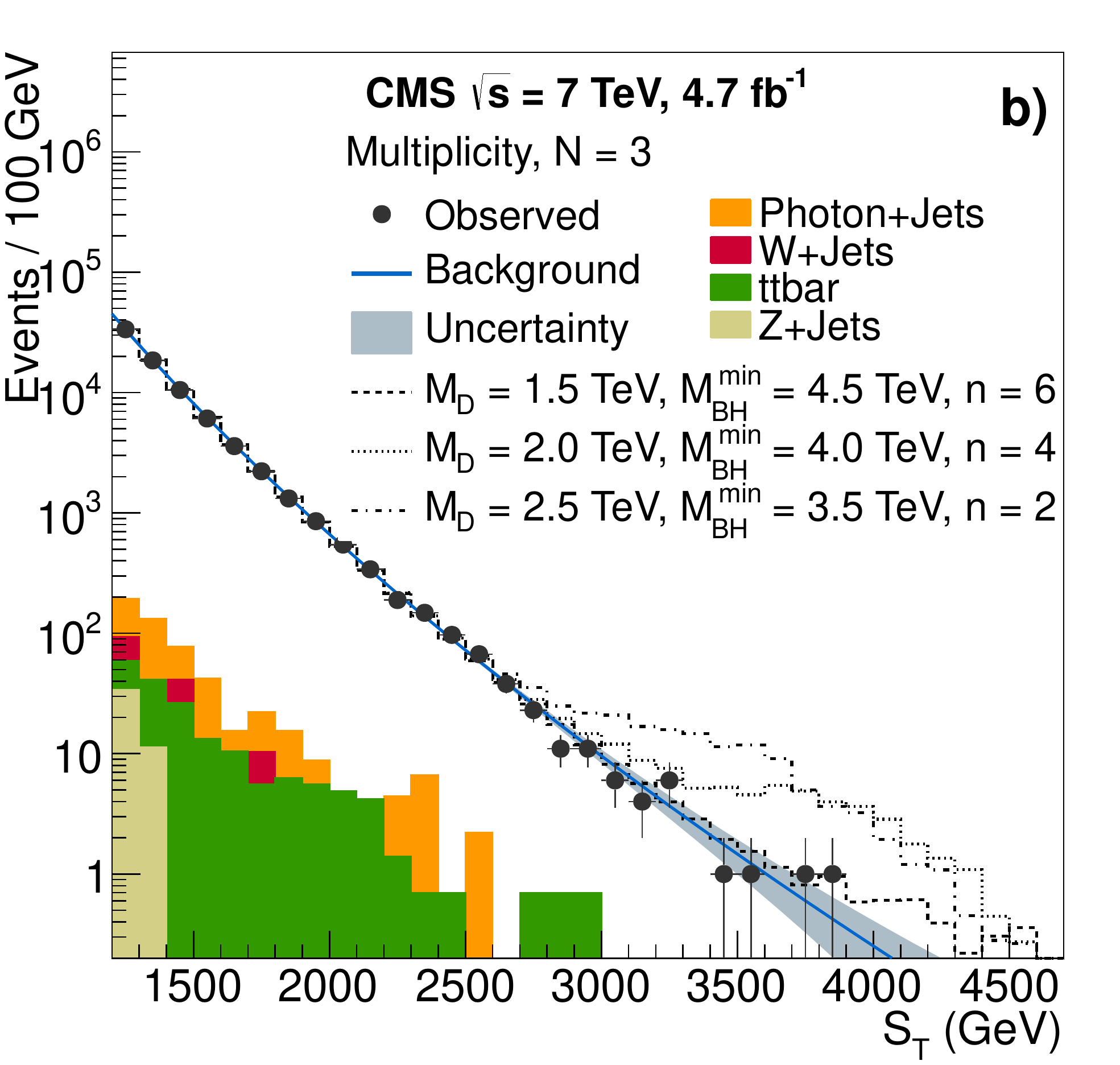}
\caption{Distribution of the total transverse energy, \ST, for low-multiplicity events with multiplicity: a) $N = 2$ and b) $N = 3$ photons, electrons, muons, or jets in the final state. Observed data are depicted as points with error bars; solid line with a shaded band is the background prediction and its systematic uncertainty. Non-QCD backgrounds are shown as filled histograms (not stacked). Also shown is the black hole signal for three parameter sets of the \BlackMax nonrotating black hole model, demonstrating that signal contamination in the fit region of $1200-2800$\GeV would be small.}
\label{fig:STN}
\end{figure}

We use low-multiplicity data with $N = 2$ and $N = 3$ to obtain the background shape by fitting the \ST distributions between $1200$ and $2800$\GeV with the ansatz function ${P_0 (1 + x)^{P_1}}/{x^{P_2 + P_3 \log(x)}}$, which is shown in Fig.~\ref{fig:STN} as a solid line. No evidence of new physics has been observed in this region in a dedicated analysis~\cite{dijets1}. To estimate the systematic uncertainty of the method, the same \ST distributions are fitted with two additional functions, ${P_0}/{(P_1 + P_2 x + x^2)^{P_3}}$ and ${P_0}/{(P_1 + x)^{P_2}}$. Thus, an envelope of functions is formed (shown as the shaded area in Fig.~\ref{fig:STN}) and is used as the systematic uncertainty.

\begin{figure}[htbp]
\centering
\includegraphics[width=0.45\textwidth]{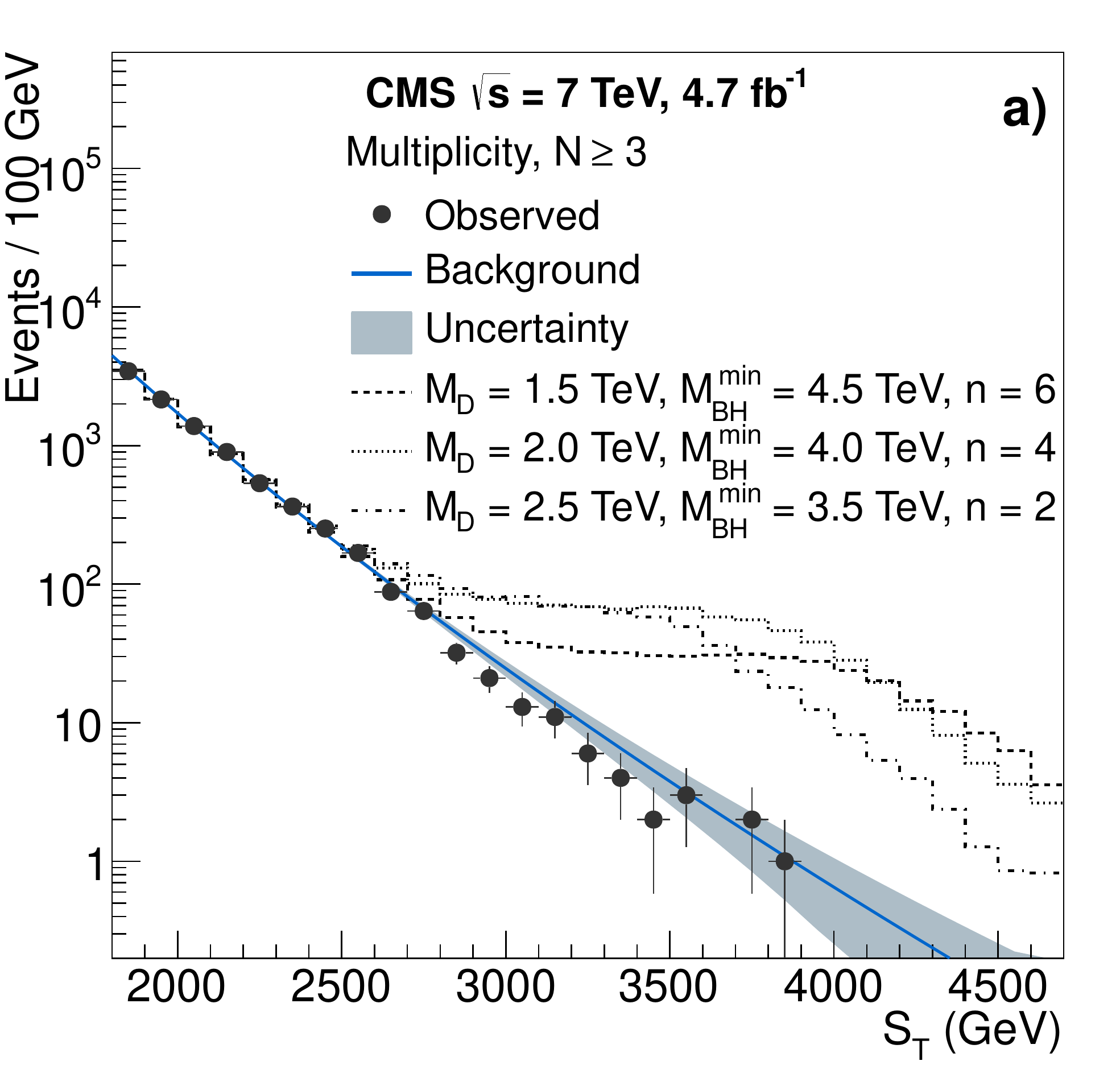}
\includegraphics[width=0.45\textwidth]{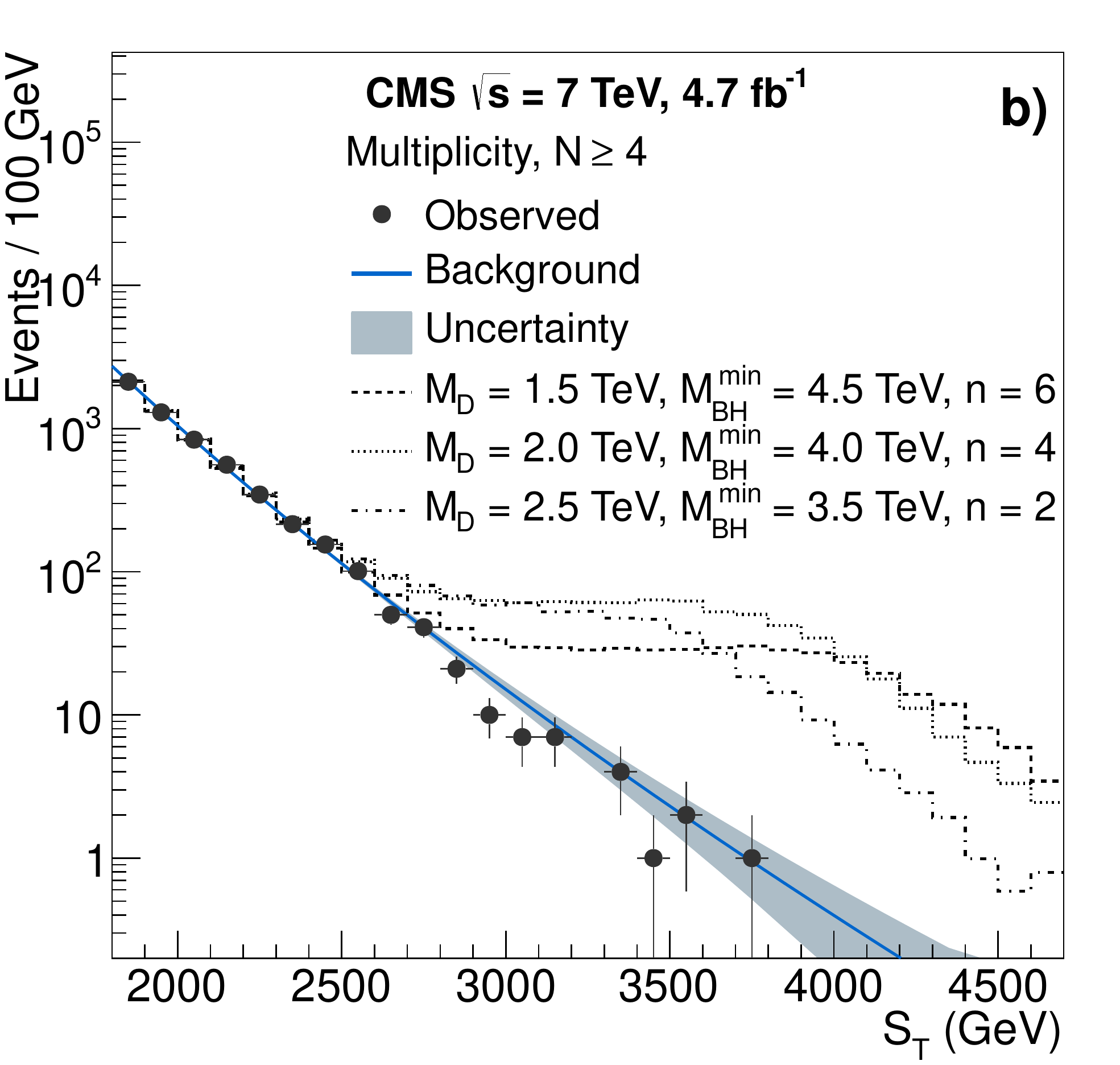}
\includegraphics[width=0.45\textwidth]{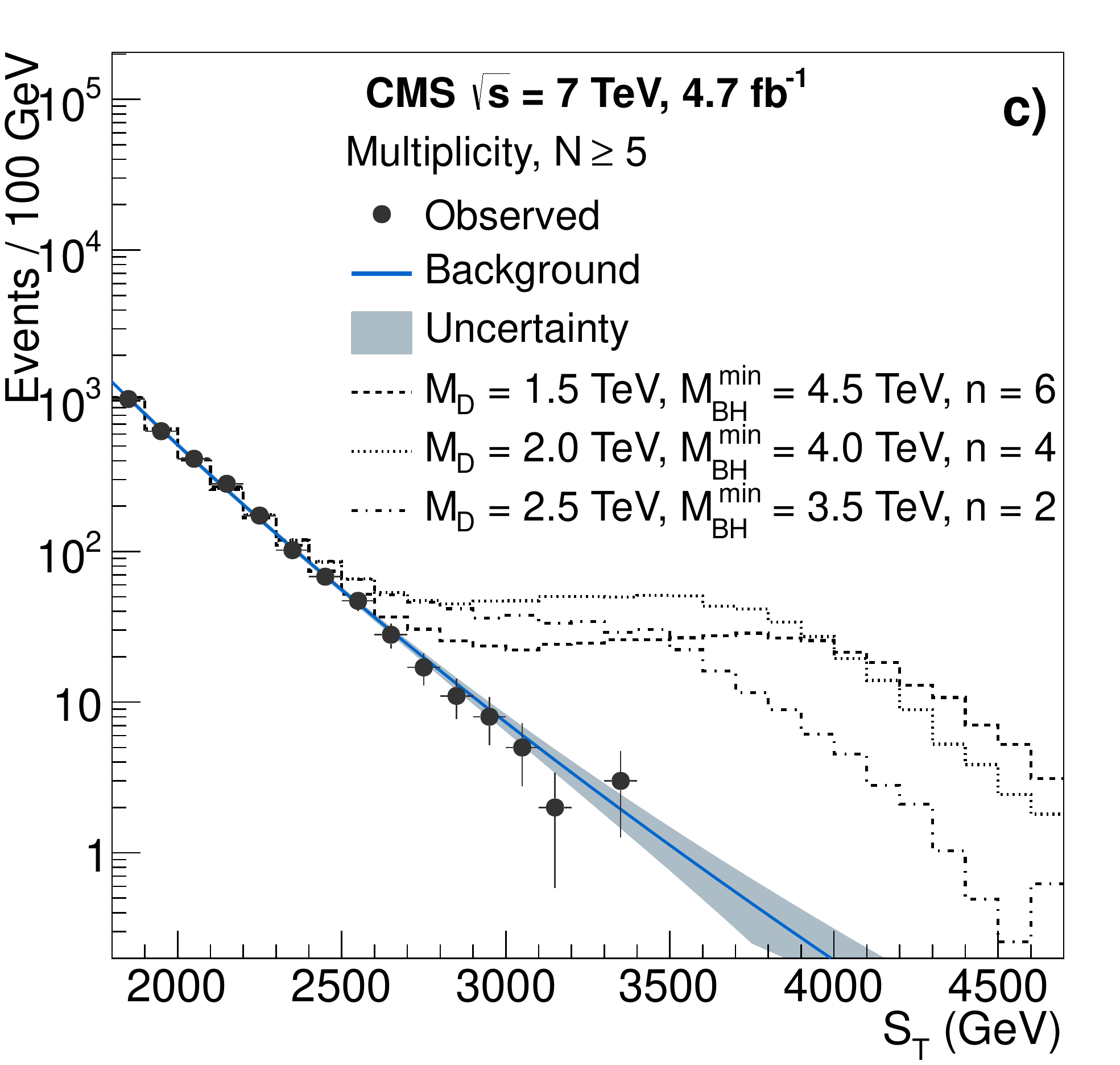}
\includegraphics[width=0.45\textwidth]{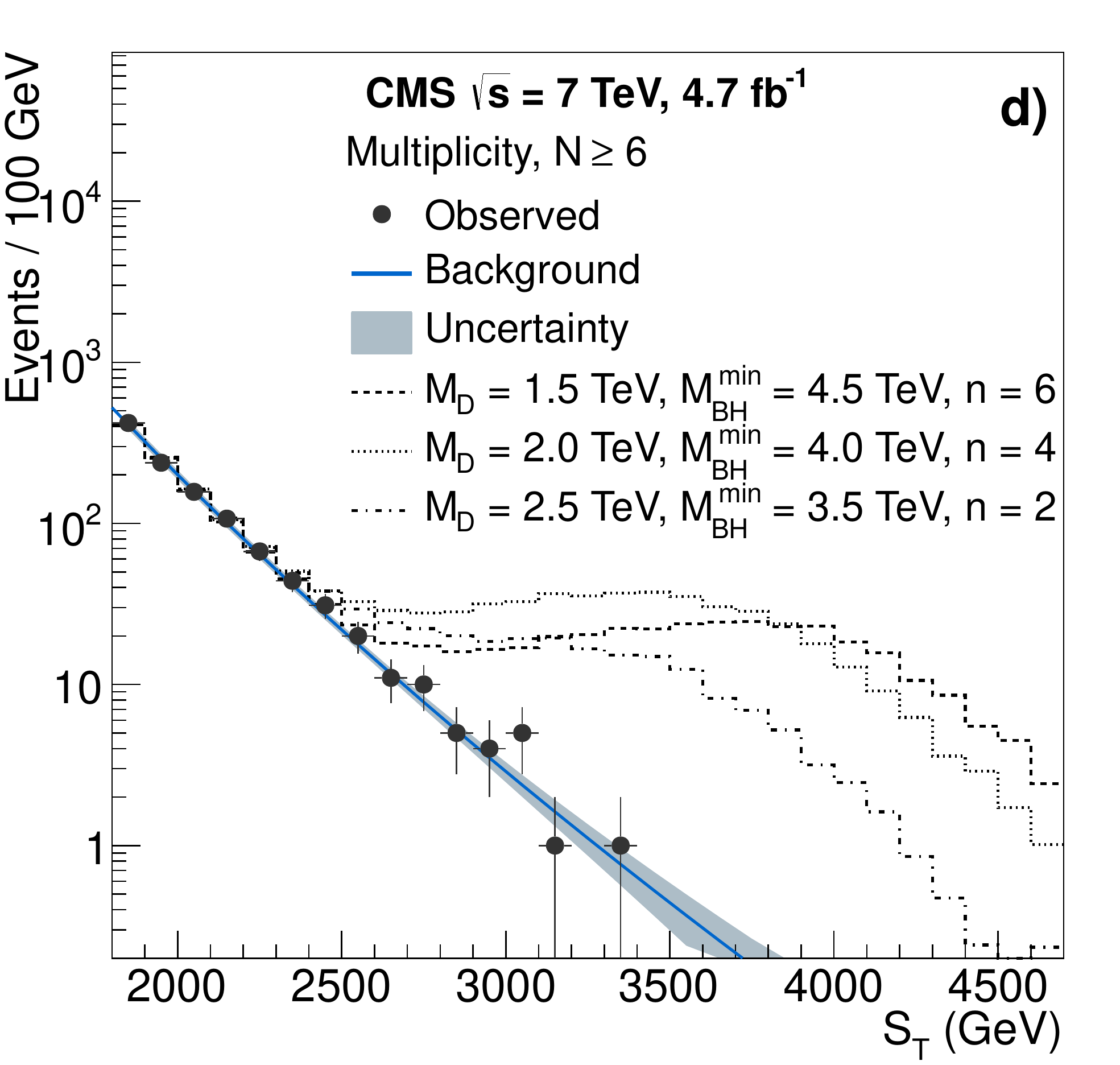}
\includegraphics[width=0.45\textwidth]{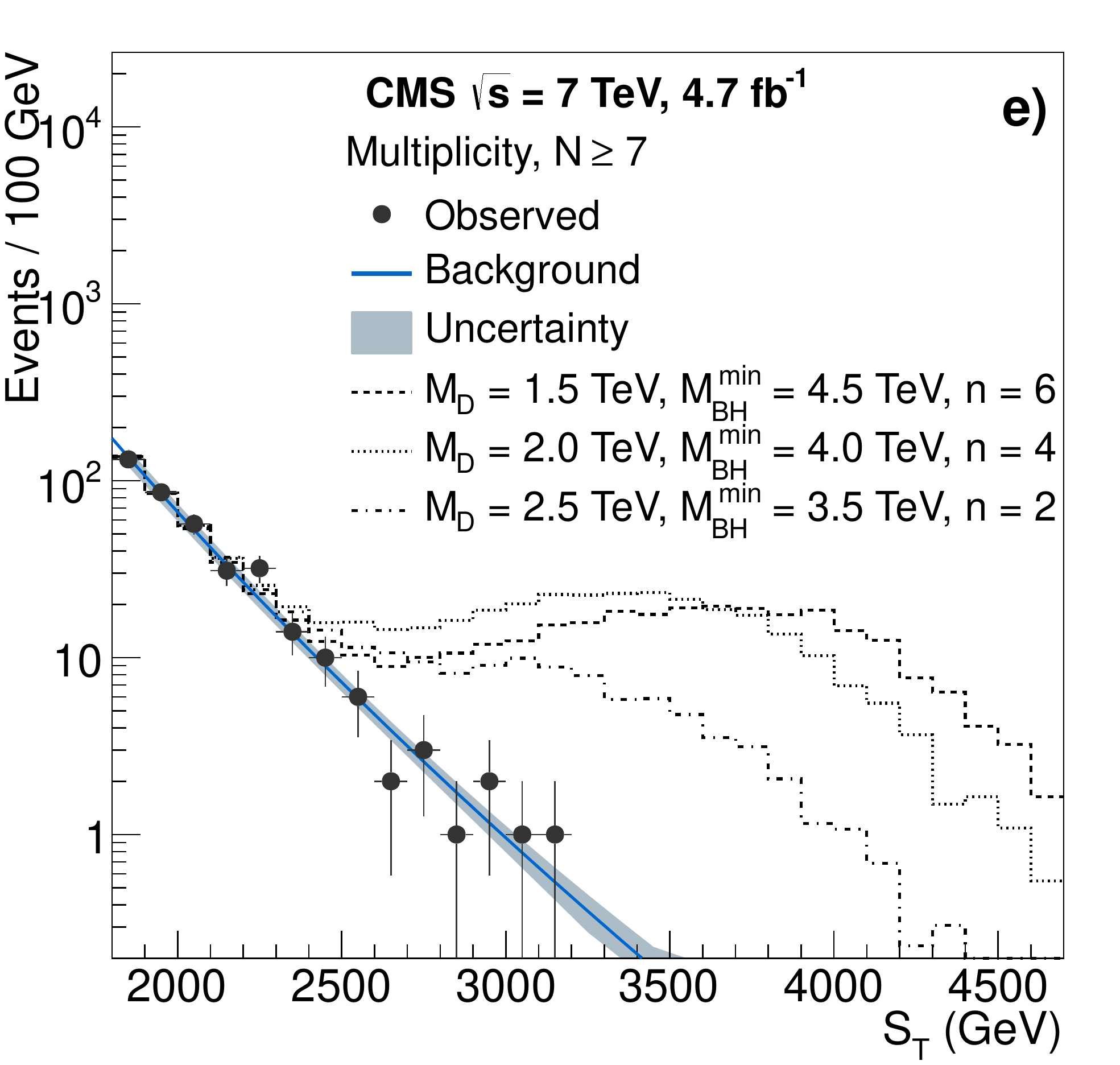}
\includegraphics[width=0.45\textwidth]{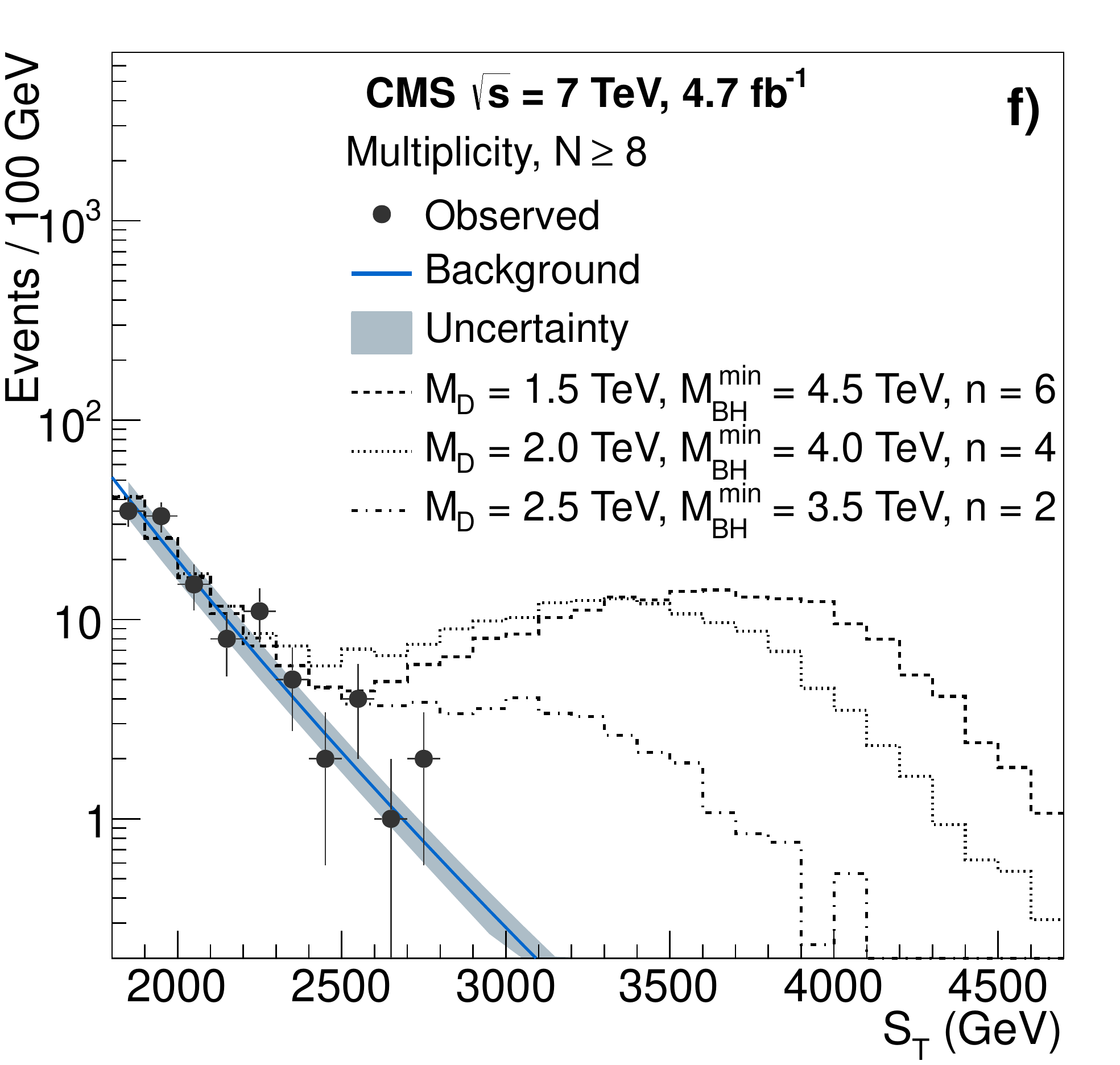}
\caption{Distribution of the total transverse energy, \ST, for events with multiplicity: a) $N \geq 3$, b) $N \geq 4$, c) $N \geq 5$, d) $N \geq 6$, e) $N \geq 7$, and f) $N \geq 8$ objects (photons, electrons, muons, or jets) in the final state. Observed data are depicted as points with error bars; the solid line with a shaded band is the background prediction and its systematic uncertainty. Also shown are the expected semiclassical black hole signals for three parameter sets of the \BlackMax nonrotating black hole model. Here, $M_\mathrm{BH}^\text{min}$ is the minimum black hole mass, \MD is the multidimensional Planck scale, and $n$ is the number of extra dimensions.}
\label{fig:STinclusive}
\end{figure}

\begin{figure}[htbp]
\centering
\includegraphics[width=0.45\textwidth]{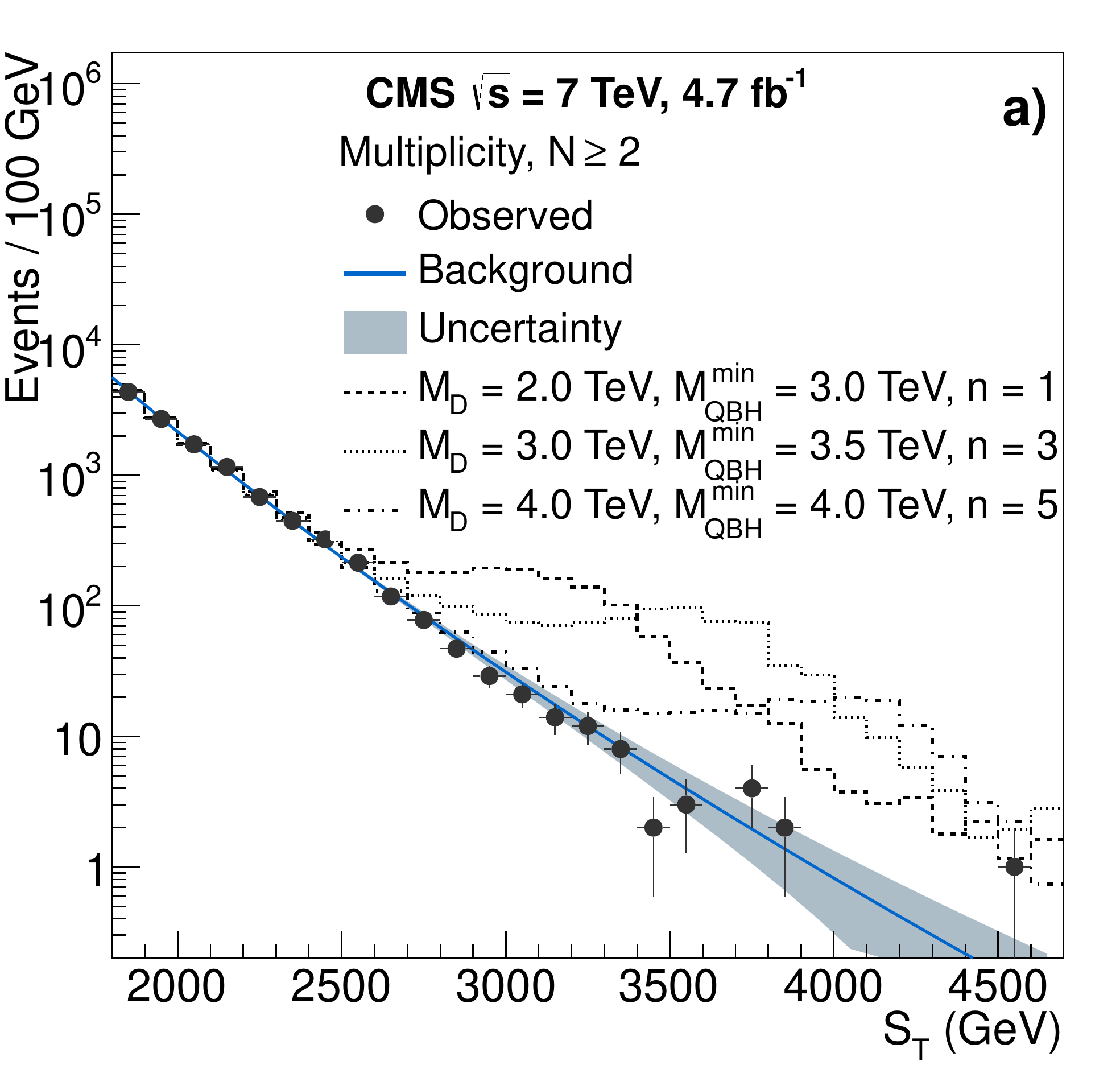}
\includegraphics[width=0.45\textwidth]{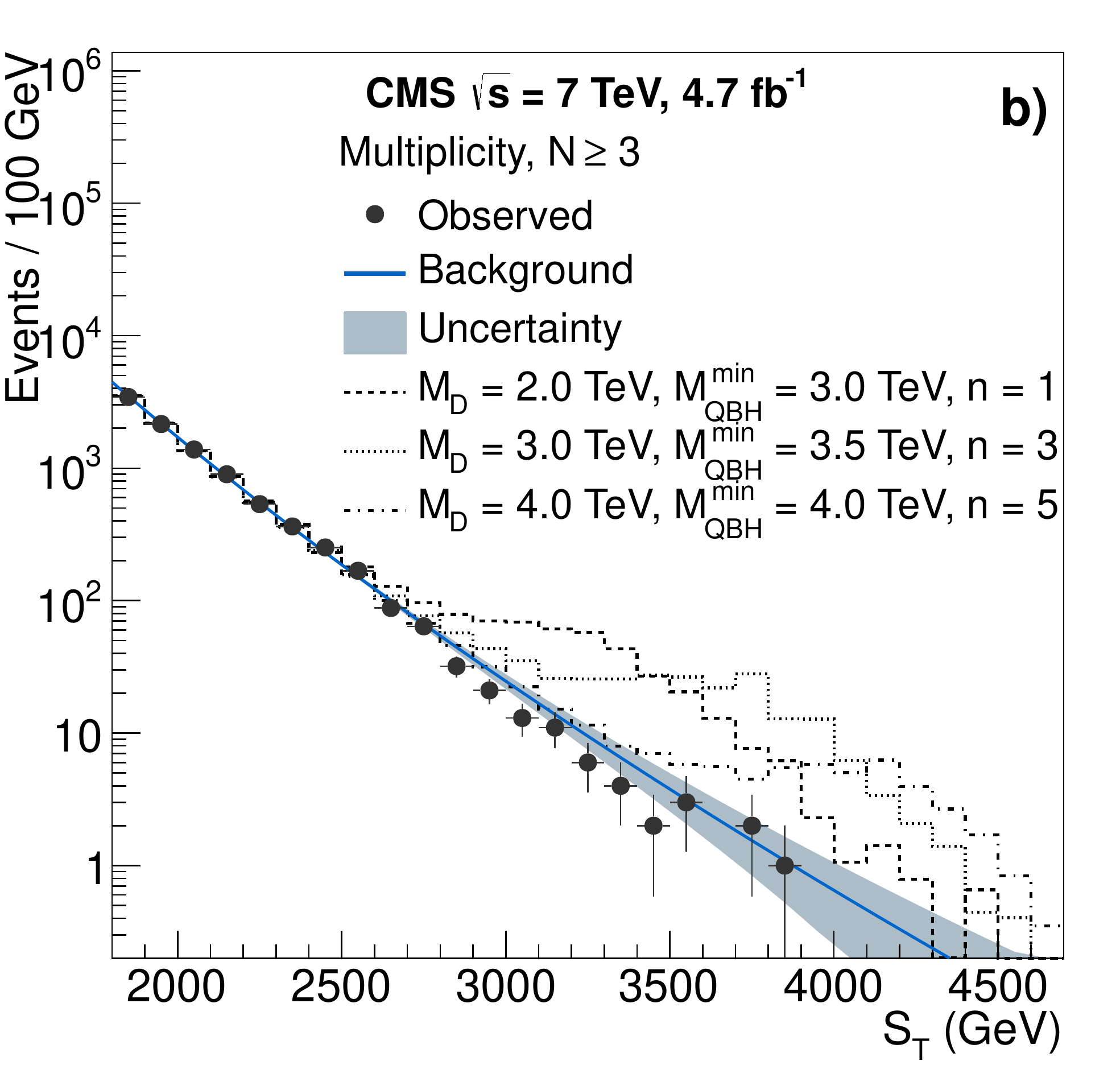}
\caption{Distribution of the total transverse energy, \ST, for events with multiplicity: a) $N \geq 2$ and b) $N \geq 3$ objects in the final state. Observed data are depicted as points with error bars; the solid line with a shaded band is the background prediction and its systematic uncertainty. Also shown are the expected quantum black hole signals for three parameter sets. Here, $M_\mathrm{QBH}^\mathrm{min}$ is the minimum quantum black hole mass, \MD is the multidimensional Planck scale, and $n$ is the number of extra dimensions.}
\label{fig:STinclusiveQBH}
\end{figure}

Figure~\ref{fig:STinclusive} shows the fit result of the background prediction for the inclusive samples with high object multiplicity events. Here, the shape of the \ST distribution obtained from the $N = 2$ sample is normalized to the observed data in the range $1800$ to $2200$\GeV, where no signal contribution is expected. Also shown are the expected semiclassical black hole signals for three parameter sets of the \BlackMax nonrotating black hole model. The results are presented separately for six different values of the minimum final state multiplicity. The data agree with the background shapes from the low-multiplicity samples and do not exhibit evidence for new physics. Figure~\ref{fig:STinclusiveQBH} shows a similar comparison of the experimental \ST distribution with the predicted signal for three parameter sets of the \textsc{qbh} quantum black hole model. In this case the comparison is shown separately for just two values of the minimum final state multiplicity, reflecting the different decay characteristics expected for quantum black holes compared to semiclassical black holes.

\section{Results}

In order to set exclusion limits on black hole production, we assign systematic uncertainties on the background estimate varying from $3\%$ to $300\%$ in the \ST range used in this search. These uncertainties are dominated by the uncertainties from using various fit ansatz functions (2\%--300\%), which are added in quadrature to the second-largest contribution, which arise from the normalization statistical uncertainty (2\%--21\%). The integrated luminosity is measured with $4.5\%$ uncertainty~\cite{lumi10,lumi} utilizing information from the forward calorimeters. The signal uncertainty is dominated by the jet energy scale uncertainty of $\approx 2\%$~\cite{JES}, which translates into $2\%$ uncertainty on the signal. An additional $2\%$ uncertainty on the signal acceptance comes from the variation of acceptance obtained with the default MSTW2008lo68cl PDF library and PDFs within the CTEQ61 and CTEQ66 error sets~\cite{CTEQ}.

Given the significant model dependence of the black hole production cross section and decay patterns, it is not practical to test all different variations of model parameters, offered by recent black hole event generators, in a dedicated search. This study considers some 700 different signal MC samples, yet it does not come close to spanning the entire parameter space of the models; scaling the number of signal samples up and presenting the results for every model therefore becomes impractical. Hence, we first present the results of our search in a generic, model-independent way, which would allow others to probe additional models using parton-level MC information, possibly augmented with a very basic detector simulation. To facilitate such an approach, we provide model-independent limits on the cross section times the acceptance for new physics production in high-\ST inclusive final states for $N \geq 3, 4, 5, 6, 7$, and $8$. The limits are set using a modified frequentist $\mathrm{CL_{s}}$ method~\cite{PDG,cls} with log-normal prior used to marginalize nuisance parameters in the likelihood function.

\begin{figure}[htbp]
\centering
   \includegraphics[height=0.40\textwidth]{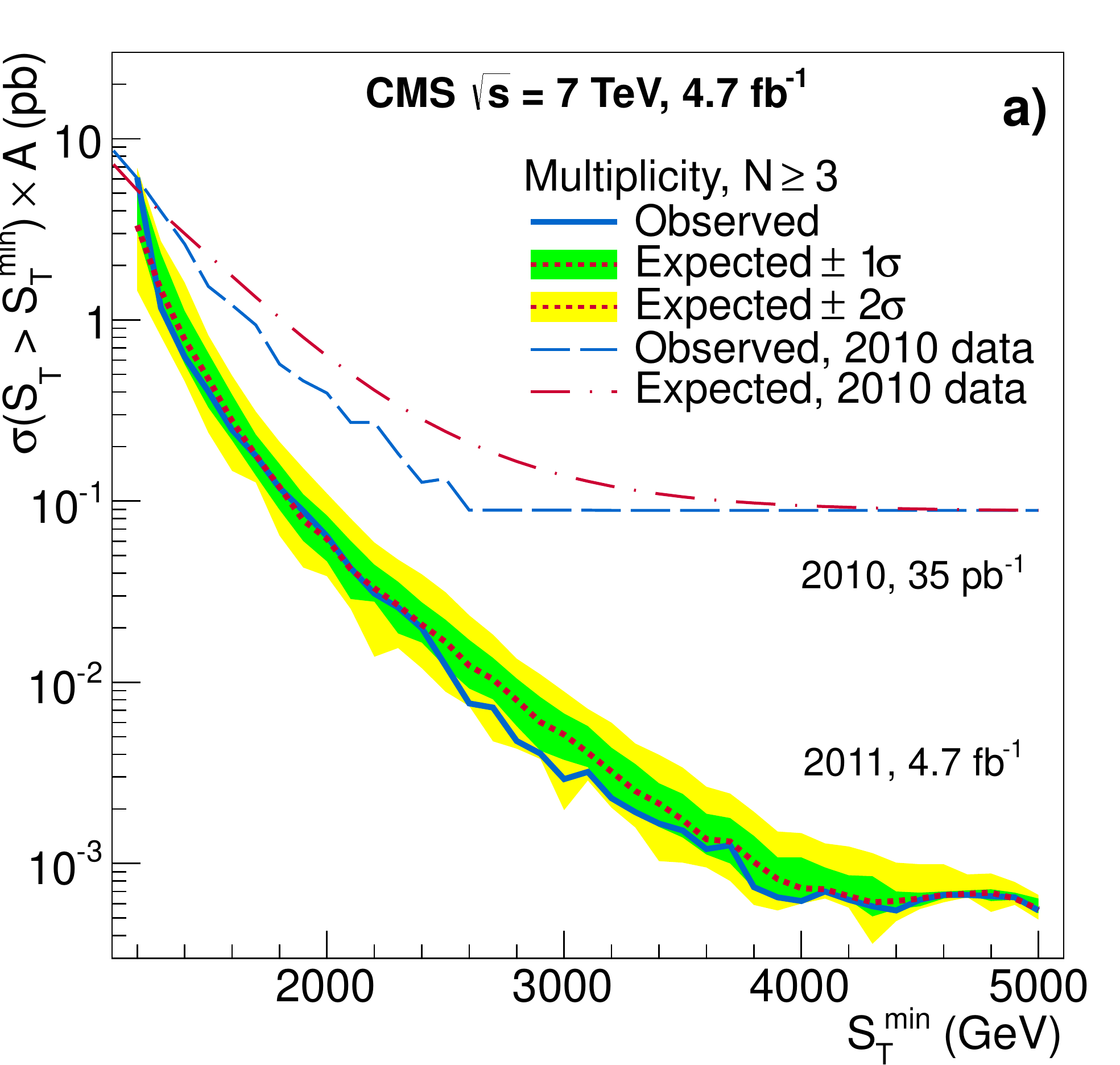}
   \includegraphics[height=0.40\textwidth]{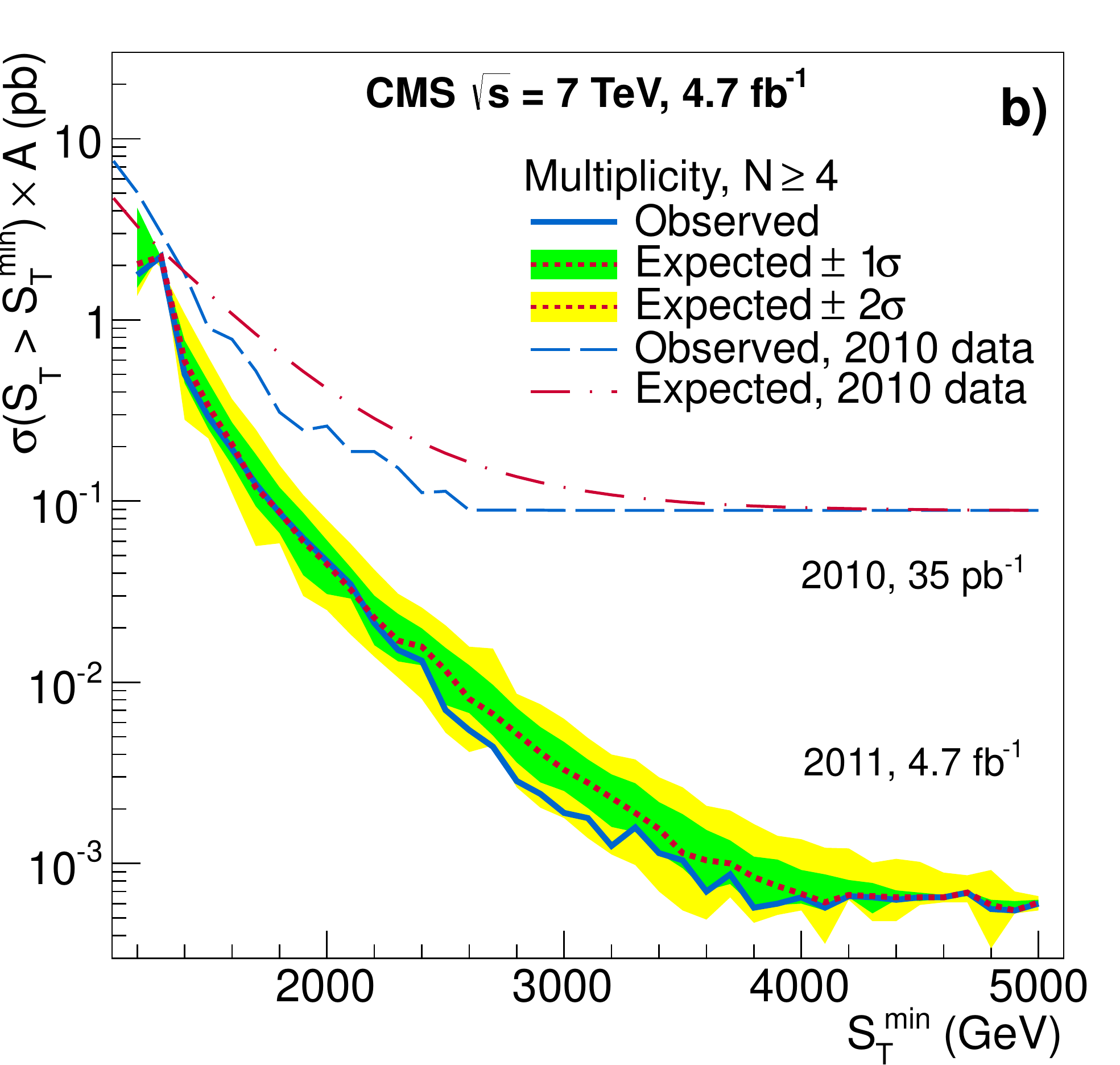}
   \includegraphics[height=0.40\textwidth]{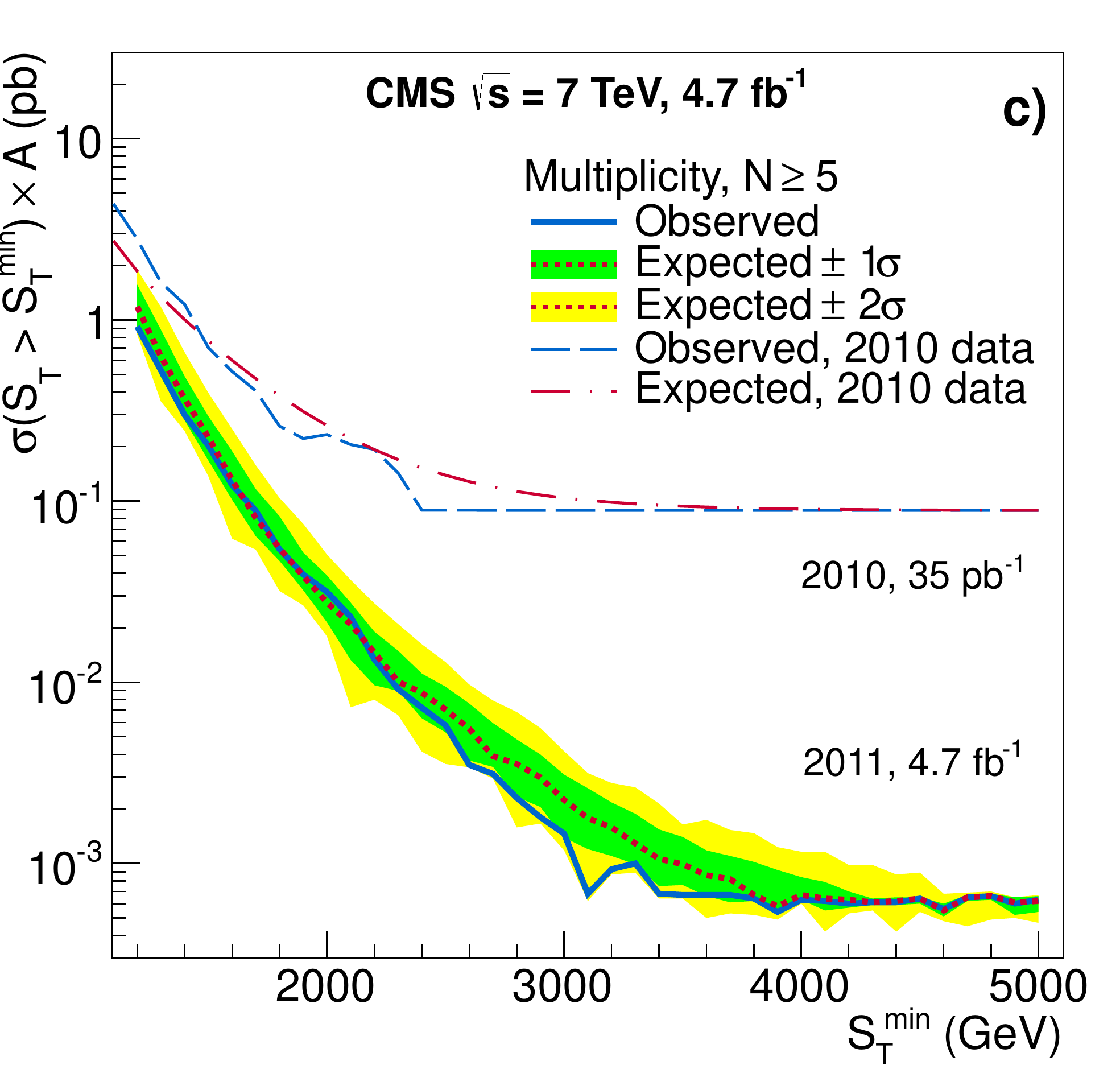}
   \includegraphics[height=0.40\textwidth]{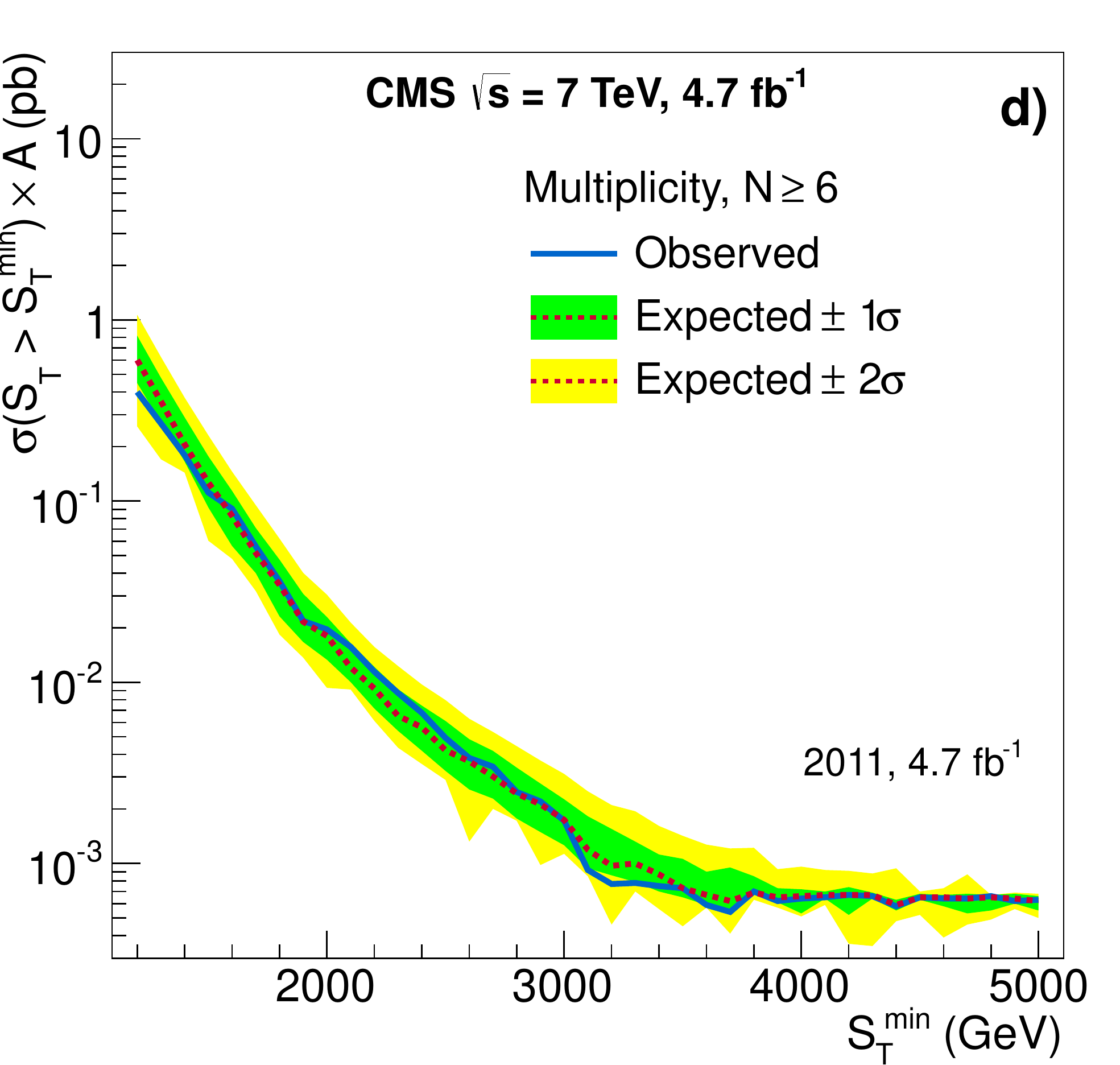}
   \includegraphics[height=0.40\textwidth]{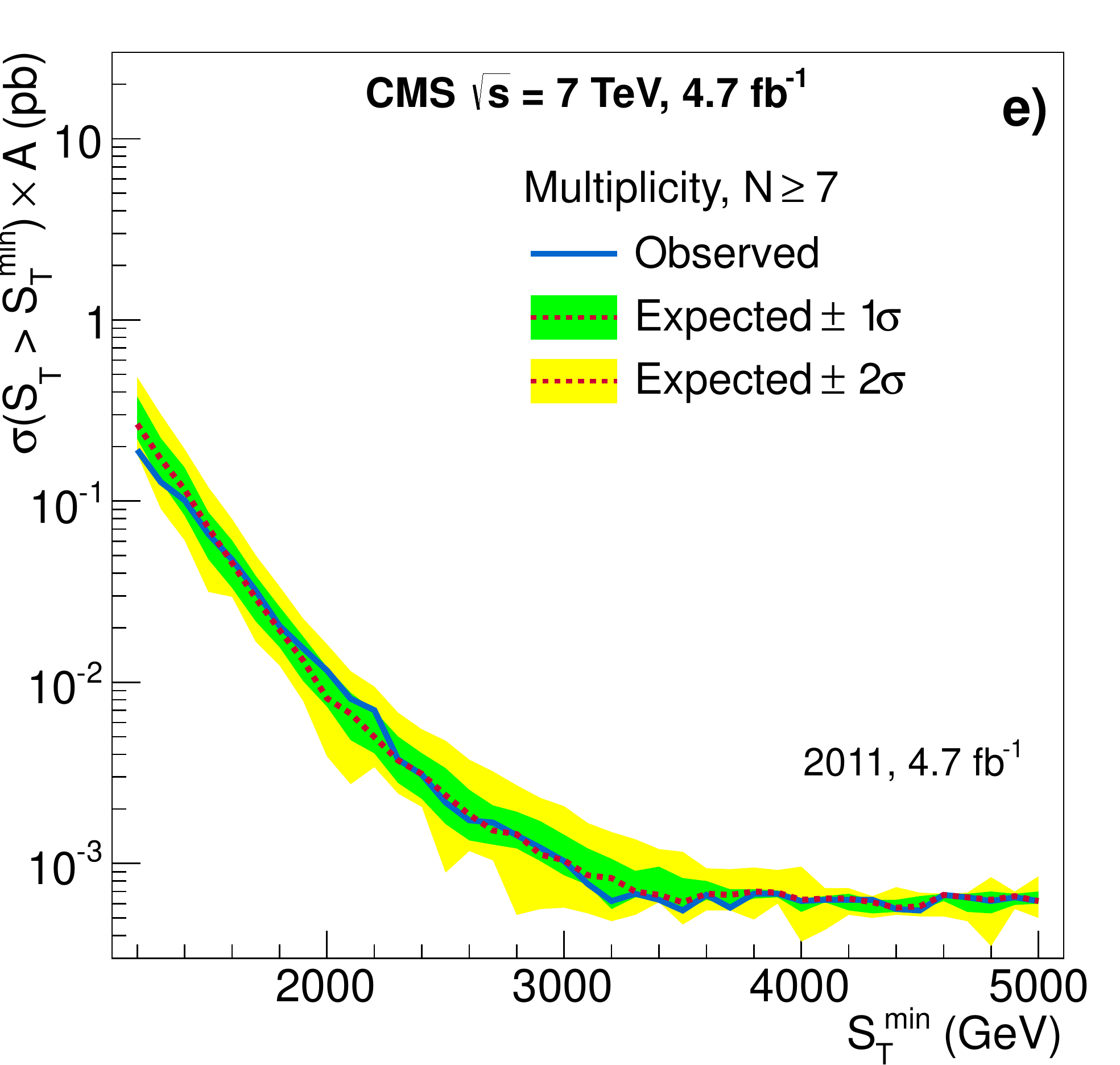}
   \includegraphics[height=0.40\textwidth]{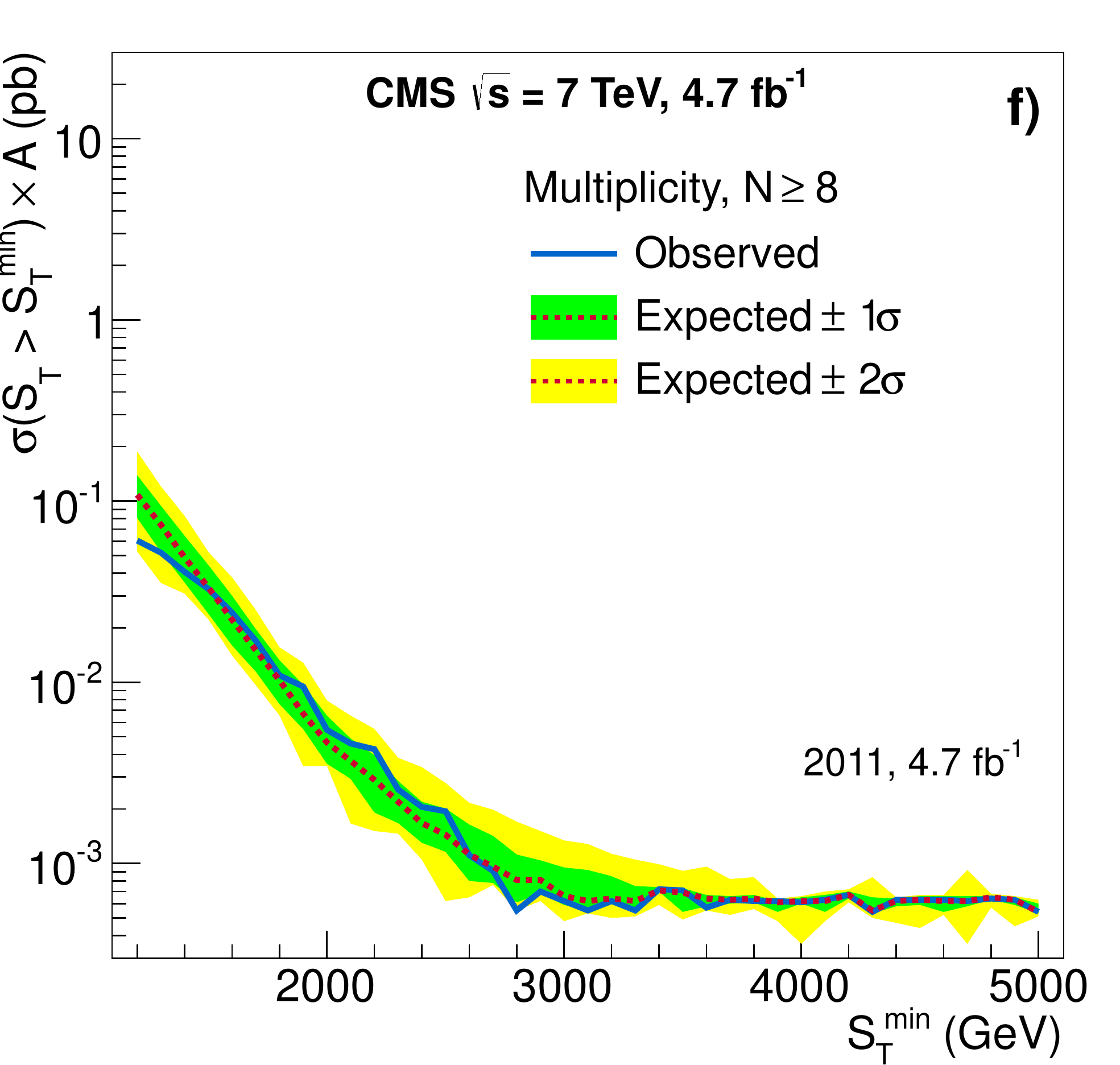}
\caption{Model-independent $95\%$ CL upper cross section limits for counting experiments with $\ST > \ST^\mathrm{min}$ as a function of $\ST^\mathrm{min}$ for events with multiplicity: a) $N \ge 3$, b) $N \ge 4$, c) $N \ge 5$, d) $N \ge 6$, e) $N \ge 7$, and f) $N \ge 8$. The blue solid (red dotted) lines correspond to an observed (expected) limit for nominal signal acceptance uncertainty of $5\%$, compared to observed (expected) limits obtained with 2010 CMS data and shown as blue dashed (red dash-dotted) line. The green and yellow bands represent one and two standard deviations from the expected limits.}
\label{fig:modind_limits}
\end{figure}

Figure~\ref{fig:modind_limits} shows $95\%$ confidence level (CL) limits from a counting experiment placed on the experimentally reconstructed value $\ST > \ST^\mathrm{min}$ as a function of $\ST^\mathrm{min}$, which can be used to test models of new physics that result in these final states, including (but not limited to) an even broader variety of black hole models than we covered in this analysis. The $95\%$ CL limits from $2010$ data~\cite{CMSBH} are also shown in Fig.~\ref{fig:modind_limits} for comparison. The present model-independent limits are roughly $0.6$\unit{fb} for high values of \ST, representing a two orders of magnitude improvement over the limits reported in our first publication~\cite{CMSBH}. Given the higher statistics of the $2011$ sample, we are also able to extend these limits to the $N \ge 6, 7$, and $8$ cases.

For a specific subset of the black hole models~\cite{CMSBH} that are being probed, we also set dedicated limits on semiclassical and quantum black hole and string-ball production performing counting experiments using optimized \ST and $N$ selections. It should be noted that the semiclassical approximation used for deriving the cross section within respective benchmark scenarios is expected to break down for many of the points probed, a point emphasized in a recent critique~\cite{Park}. Thus, these limits should be treated as indicative, rather than precise.

\begin{figure}[htbp]
\centering
\includegraphics[width=0.6\textwidth]{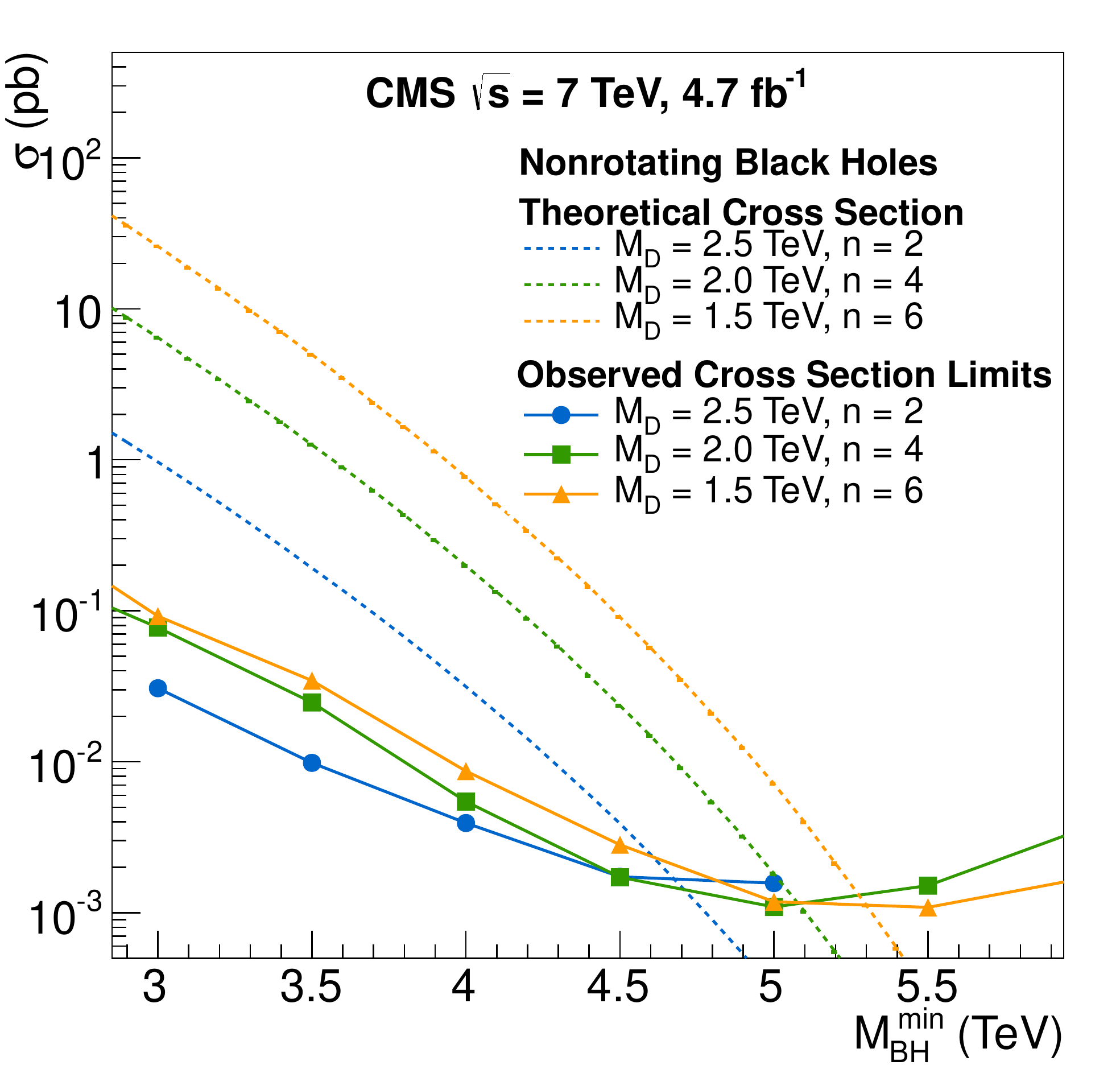}
\caption{Cross section limits at $95\%$ CL from the counting experiments optimized for various black hole parameter sets (solid lines) compared with signal production cross sections from the \BlackMax generator (dashed lines) as a function of minimum black hole mass.}
\label{fig:BHL}
\end{figure}

The signal ($S$) significance is optimized in the presence of background ($B$) using a test statistic $S/\sqrt{S+B}$ for each set of model parameters. The optimum choices of \ST and $N$ for a few illustrative benchmark scenarios are listed in Table~\ref{tab:BH}, as well as the predicted number of background events, expected number of signal events, and the observed number of events in data. The corresponding cross section limits at the 95\% CL from the counting experiments are shown in Fig.~\ref{fig:BHL} for a small subset of characteristic signal points.

\begin{table}[htbp]
\begin{center}
\caption{Details of some of the \BlackMax nonrotating black hole model parameters probed in the analysis. Shown are: multidimensional Planck scale (\MD), minimum black hole mass ($M_\mathrm{BH}$), number of extra dimensions ($n$), corresponding leading order cross sections ($\sigma$), and optimal selections on the minimum decay multiplicity ($N \geq N^{\min}$) and minimum \ST, as well as signal acceptance (A), expected number of signal events ($N^\text{sig}$), number of observed events in data ($N^\text{data}$), expected background ($N^\text{bkg}$) with its systematic uncertainty, and observed ($\sigma^{95}$) and expected ($\langle\sigma^{95}\rangle$) limit on the signal at $95\%$ CL.}
\medskip
\label{tab:BH}
\rm\footnotesize\rm
\begin{tabular}{|c|c|c|c|c|c|c|c|c|c|c|c|}
\hline
\MD & $M_\mathrm{BH}$ & $n$ & $\sigma$ & $N^\mathrm{min}$ & $\ST^\text{min}$ & A & $N^\text{sig}$ & $N^\text{data}$ & $N^\text{bkg}$ & $\sigma^{95}$ & $\langle\sigma^{95}\rangle$ \\
(TeV) & (TeV) & & (pb) & & (TeV) & (\%) & & & & (pb) & (pb)\\
\hline
1.5 & 3.0 & 6 & 26 & 3 & 1.9 & 88.0 & 110000 & 5999 & $5970 \pm 180$ & 0.092 & 0.091\\
1.5 & 3.5 & 6 & 5.0 & 3 & 2.2 & 88.0 & 21000 & 1565 & $1590 \pm 66$ & 0.035 & 0.06\\
1.5 & 4.0 & 6 & 0.77 & 4 & 2.5 & 85.8 & 3100 & 245 & $280 \pm 24$ & 0.0087 & 0.014\\
1.5 & 4.5 & 6 & 0.091 & 5 & 2.8 & 81.8 & 350 & 29 & $42 \pm 7$ & 0.0028 & 0.0044\\
1.5 & 5.0 & 6 & 0.0071 & 6 & 3.2 & 73.5 & 25 & 1 & $3.7 \pm 1.3$ & 0.0012 & 0.0015\\
1.5 & 5.5 & 6 & 0.0003 & 7 & 3.7 & 61.8 & 0.82 & 0 & $0.21 \pm 0.14$ & 0.0011 & 0.0011\\
2.0 & 3.0 & 4 & 6.5 & 3 & 2.0 & 82.5 & 25000 & 3847 & $3810 \pm 120$ & 0.077 & 0.075\\
2.0 & 3.5 & 4 & 1.3 & 3 & 2.4 & 79.7 & 4700 & 667 & $690 \pm 45$ & 0.025 & 0.026\\
2.0 & 4.0 & 4 & 0.20 & 3 & 2.8 & 77.1 & 720 & 95 & $140 \pm 23$ & 0.0054 & 0.0094\\
2.0 & 4.5 & 4 & 0.024 & 4 & 3.2 & 69.5 & 77 & 8 & $19 \pm 6$ & 0.0017 & 0.0032\\
2.0 & 5.0 & 4 & 0.0018 & 5 & 3.6 & 61.3 & 5.2 & 0 & $2.3 \pm 1.3$ & 0.0011 & 0.0014\\
2.5 & 3.0 & 2 & 0.97 & 3 & 2.4 & 62.3 & 2800 & 667 & $690 \pm 45$ & 0.031 & 0.034\\
2.5 & 3.5 & 2 & 0.19 & 3 & 2.7 & 65.3 & 590 & 159 & $210 \pm 28$ & 0.0098 & 0.016\\
2.5 & 4.0 & 2 & 0.031 & 3 & 3.2 & 57.7 & 85 & 18 & $31 \pm 11$ & 0.0039 & 0.0054\\
2.5 & 4.5 & 2 & 0.0039 & 4 & 3.6 & 46.7 & 8.5 & 1 & $4.6 \pm 2.7$ & 0.0017 & 0.0024\\
2.5 & 5.0 & 2 & 0.0003 & 4 & 4.1 & 41.3 & 0.59 & 0 & $0.86 ^{+0.89}_{-0.86}$ & 0.0016 & 0.0017\\

\hline
\end{tabular}
\end{center}
\end{table}

\begin{figure}[htbp]
\centering
\includegraphics[width=0.45\textwidth]{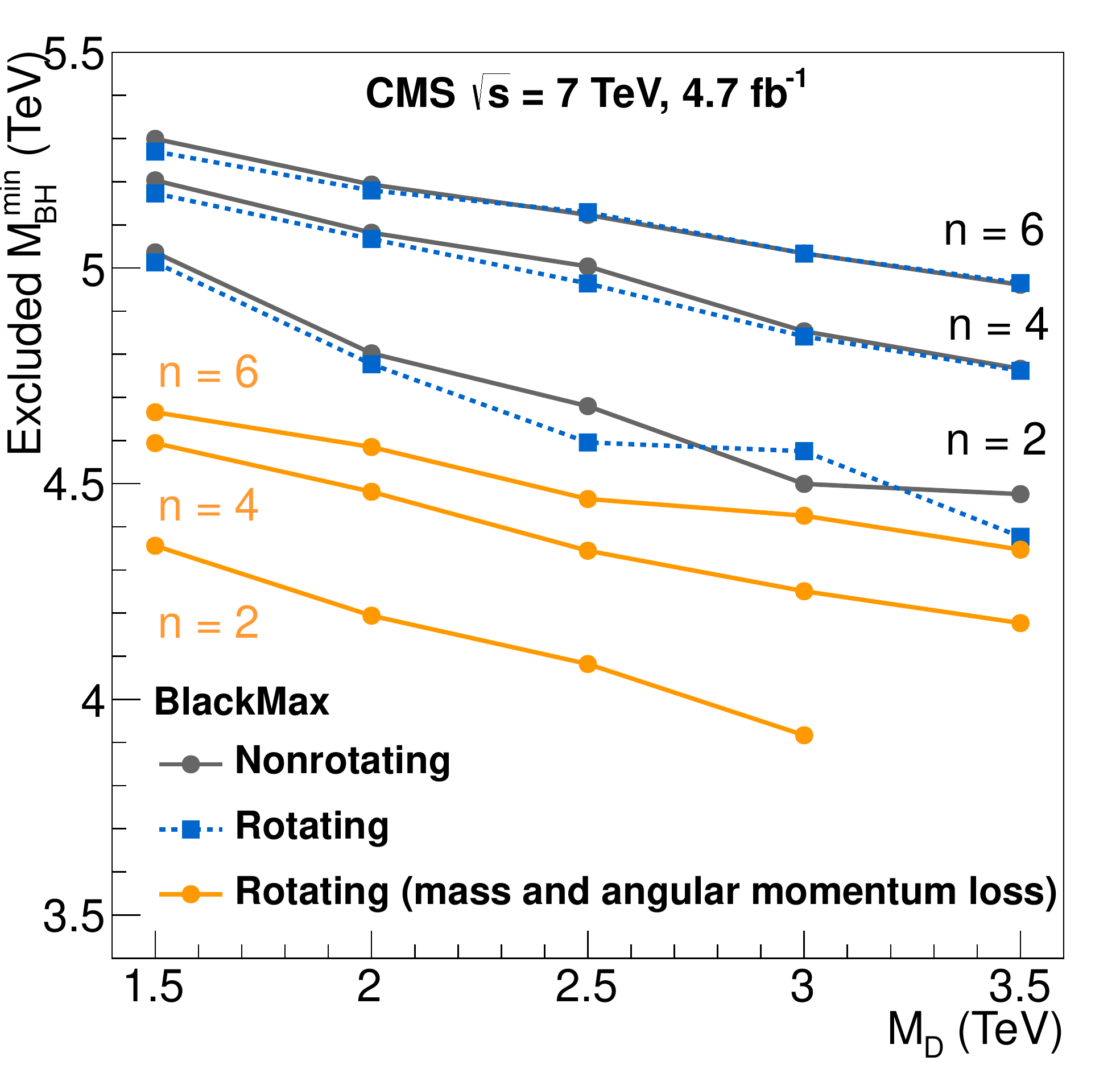}
\includegraphics[width=0.45\textwidth]{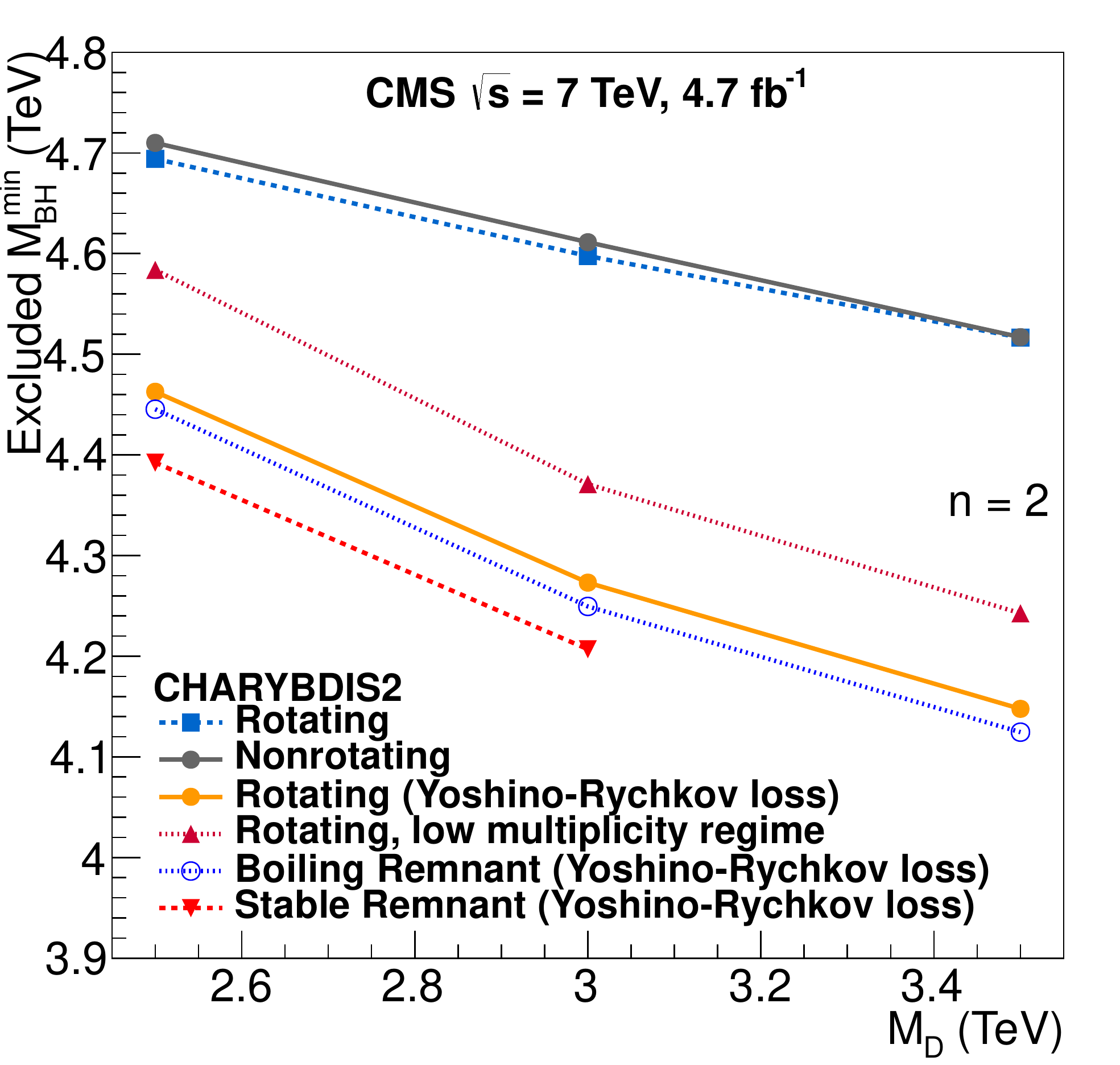}
\includegraphics[width=0.45\textwidth]{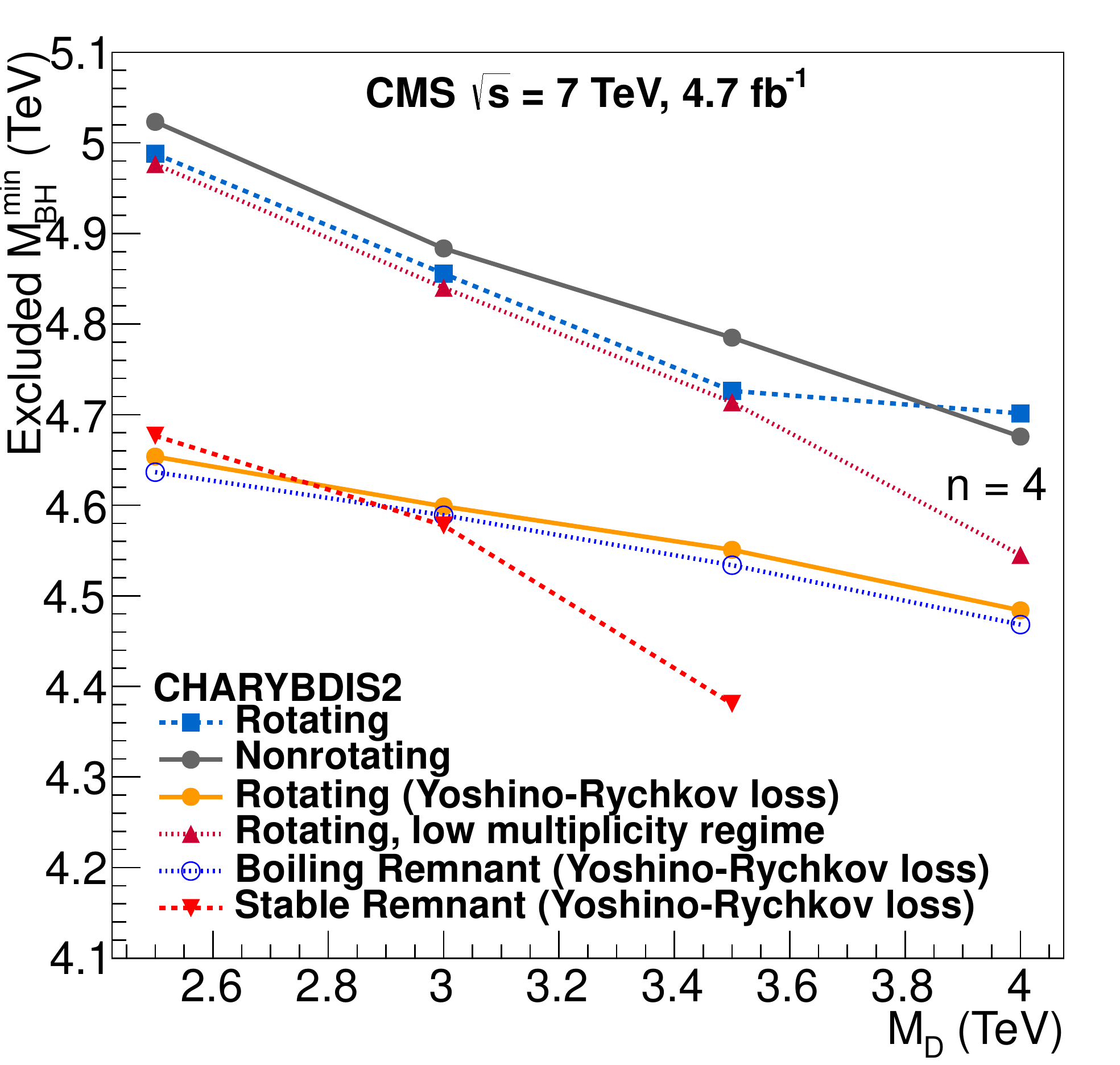}
\includegraphics[width=0.45\textwidth]{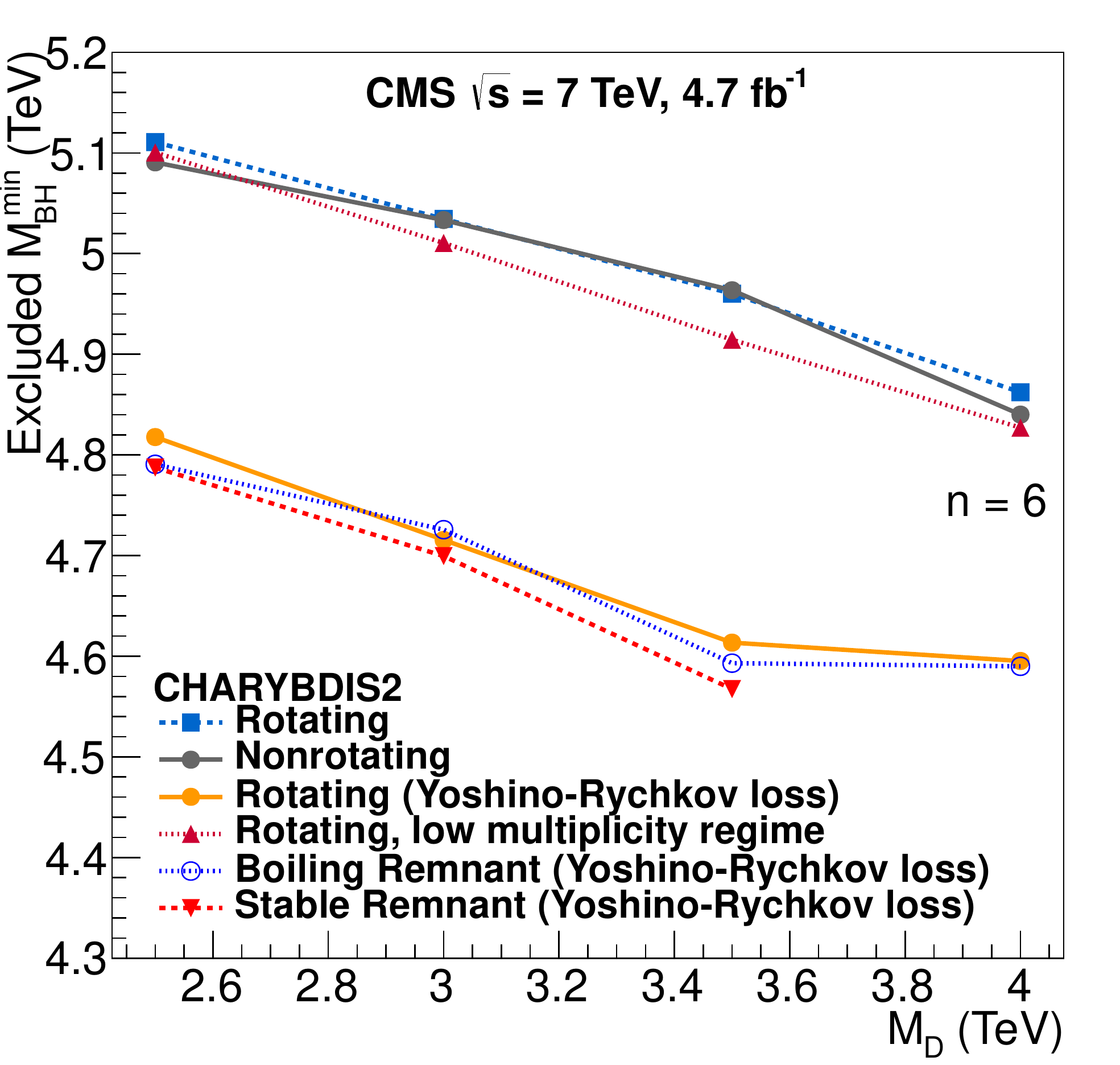}
\caption{Minimum black hole mass excluded at $95\%$ CL as function of the reduced Planck scale for various \BlackMax black hole models without the stable remnant and number of extra dimensions of two, four, and six (Top left). The minimum black hole mass, excluded at $95\%$ CL, as function of the reduced Planck scale for various \Charybdis black hole models with or without the stable remnant and number of extra dimensions $n$ of (top right) two, (bottom left) four, and (bottom right) six. The areas below each curve are excluded by this search.}
\label{fig:BHpar}
\end{figure}

\begin{figure}[htbp]
\centering
\includegraphics[width=0.6\textwidth]{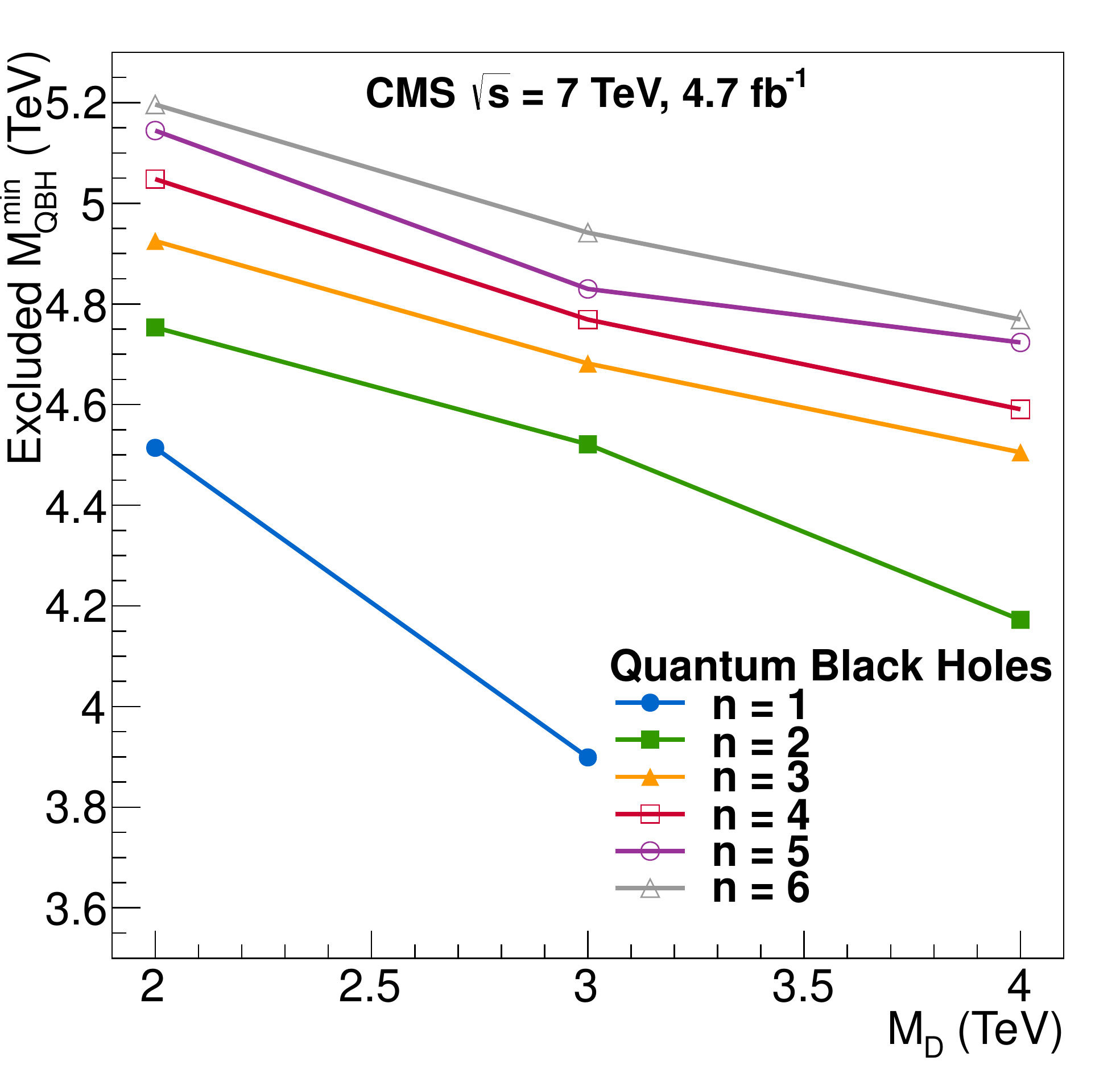}
\caption{95\% CL excluded minimum quantum black hole mass as function of the reduced Planck scale for number of extra dimensions $n$ of one (Randall--Sundrum model) and two to six (ADD model). This search excludes the areas below each curve.}
\label{fig:QBHpar}
\end{figure}

\begin{figure}[htbp]
\centering
\includegraphics[width=0.6\textwidth]{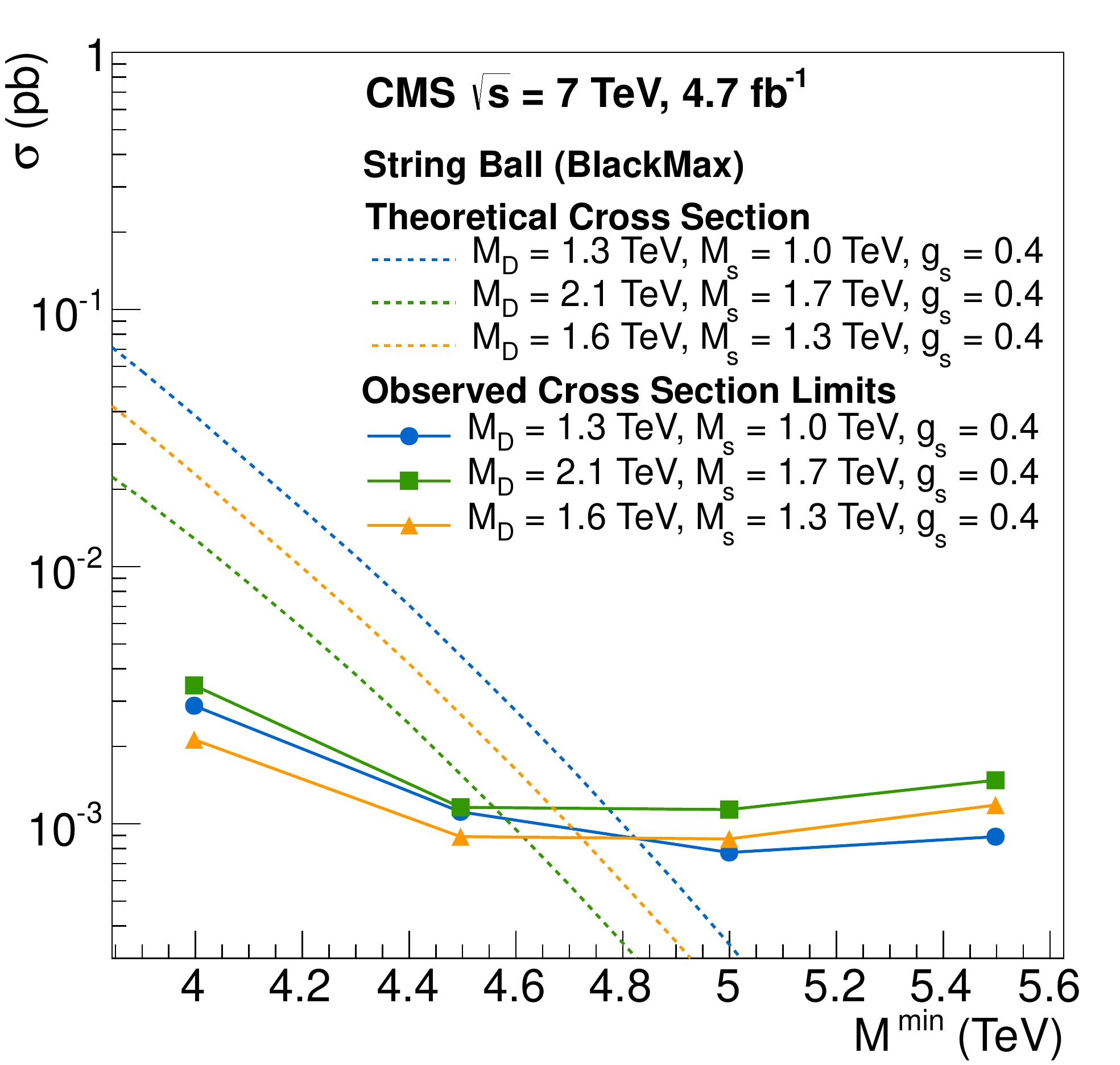}
\caption{Cross section limits at $95\%$ CL from the counting experiments optimized for various string-ball parameter sets (solid lines) compared with signal production cross section (dashed lines) as a function of minimum string-ball mass. Here, \MD is the multidimensional Planck scale, $M_S$ is the string scale, and $g_S$ is the string coupling.}
\label{fig:SB}
\end{figure}

By translating the cross section limits into ADD model expectations, we can exclude the production of semiclassical black holes with a minimum mass varying from $3.9$ to $5.3$\TeV for values of the multidimensional Planck scale $\MD \le 4$\TeV and a number of extra dimensions $n \le 6$ at $95\%$ CL (see Fig.~\ref{fig:BHpar}). The excluded minimum masses of quantum black holes are in the 3.8--5.2\TeV range for \MD up to 4\TeV and are shown in Fig.~\ref{fig:QBHpar}. The $95\%$ CL limits on the string-ball production cross section as a function of the minimum mass of the string ball are shown in Fig.~\ref{fig:SB}. We exclude string-ball production with a minimum mass from 4.6 to 4.8\TeV, depending on the model. Despite the caveats mentioned above, we consider it useful to present these results, since at present there are no alternative quantitative calculations in the regime where the semiclassical approximation breaks down. Furthermore, although the predicted cross section is very sensitive to changes in the model of black hole production and decay, varying the model assumptions results in only moderate changes to the mass limit, because of the exponential dependence of the cross section on the black hole mass. Nontheless, we emphasize that the model-independent limits set in this Letter should be used in the regime when the semiclassical approximation fails in order to obtain more reliable predictions.

\section{Conclusions}

An update has been presented of an earlier dedicated search for black holes at the LHC~\cite{CMSBH} and new model-independent limits have been set on the production of energetic multiparticle final states, which can be used to constrain a large variety of models of new physics. For the benchmark models, the following can be concluded from Fig.~\ref{fig:BHpar}: (i) since the excluded masses of rotating and nonrotating black holes are similar, the effect of black hole spin on the sensitivity of the search is small; (ii) in case of energy/momentum loss due to gravitational radiation not trapped by the forming event horizon, the excluded black hole masses are ${\sim}10\%$ lower than in the case of no losses; (iii) the choice of \ST as a discriminating variable makes the results largely insensitive to the details of the last stage of black holes evaporation, whether a stable remnant is formed or not. Numerically, the limits on the minimum semiclassical and quantum black hole and string-ball masses are in the range $3.8$ to $5.3$\TeV for a wide range of model parameters. These are the most restrictive limits on black hole production set at hadron colliders to date. Further extension of this search will be possible when the LHC energy is increased in the future.

\section*{Acknowledgements}

We congratulate our colleagues in the CERN accelerator departments for the excellent performance of the LHC machine. We thank the technical and administrative staff at CERN and other CMS institutes, and acknowledge support from: FMSR (Austria); FNRS and FWO (Belgium); CNPq, CAPES, FAPERJ, and FAPESP (Brazil); MES (Bulgaria); CERN; CAS, MoST, and NSFC (China); COLCIENCIAS (Colombia); MSES (Croatia); RPF (Cyprus); MoER, SF0690030s09 and ERDF (Estonia); Academy of Finland, MEC, and HIP (Finland); CEA and CNRS/IN2P3 (France); BMBF, DFG, and HGF (Germany); GSRT (Greece); OTKA and NKTH (Hungary); DAE and DST (India); IPM (Iran); SFI (Ireland); INFN (Italy); NRF and WCU (Korea); LAS (Lithuania); CINVESTAV, CONACYT, SEP, and UASLP-FAI (Mexico); MSI (New Zealand); PAEC (Pakistan); MSHE and NSC (Poland); FCT (Portugal); JINR (Armenia, Belarus, Georgia, Ukraine, Uzbekistan); MON, RosAtom, RAS and RFBR (Russia); MSTD (Serbia); MICINN and CPAN (Spain); Swiss Funding Agencies (Switzerland); NSC (Taipei); TUBITAK and TAEK (Turkey); STFC (United Kingdom); DOE and NSF (USA).

\bibliography{auto_generated}   
\cleardoublepage \appendix\section{The CMS Collaboration \label{app:collab}}\begin{sloppypar}\hyphenpenalty=5000\widowpenalty=500\clubpenalty=5000\textbf{Yerevan Physics Institute,  Yerevan,  Armenia}\\*[0pt]
S.~Chatrchyan, V.~Khachatryan, A.M.~Sirunyan, A.~Tumasyan
\vskip\cmsinstskip
\textbf{Institut f\"{u}r Hochenergiephysik der OeAW,  Wien,  Austria}\\*[0pt]
W.~Adam, T.~Bergauer, M.~Dragicevic, J.~Er\"{o}, C.~Fabjan, M.~Friedl, R.~Fr\"{u}hwirth, V.M.~Ghete, J.~Hammer\cmsAuthorMark{1}, M.~Hoch, N.~H\"{o}rmann, J.~Hrubec, M.~Jeitler, W.~Kiesenhofer, M.~Krammer, D.~Liko, I.~Mikulec, M.~Pernicka$^{\textrm{\dag}}$, B.~Rahbaran, C.~Rohringer, H.~Rohringer, R.~Sch\"{o}fbeck, J.~Strauss, A.~Taurok, F.~Teischinger, P.~Wagner, W.~Waltenberger, G.~Walzel, E.~Widl, C.-E.~Wulz
% ...
\vskip\cmsinstskip
\dag:~Deceased\\
1:~~Also at CERN, European Organization for Nuclear Research, Geneva, Switzerland\\
% ...

\end{sloppypar}
\end{document}